\documentclass[a4paper,11pt]{article}
\usepackage{jheppub}
\usepackage{amssymb,amsmath}
\usepackage[british]{babel}

\def\Cset{\mathbb{C}}
\def\openone{\hbox{\upshape \small1\kern-3.3pt\normalsize1}}
\newcommand{\ltsim}{\raisebox{-.6ex}{$\stackrel{\textstyle{<}}{\sim}$}}
\newcommand{\norm}[1]{\left\lVert #1\right\rVert}

\title{The Shape of Covariantly Smeared Sources in Lattice QCD}

\author[a]{Georg M. von Hippel,}
\author[a,b]{Benjamin J\"ager,}
\author[a]{Thomas D. Rae}
\author[a,b]{and Hartmut Wittig}

\affiliation[a]{PRISMA Cluster of Excellence and Institut f\"ur Kernphysik, University of Mainz, 55099 Mainz, Germany}
\affiliation[b]{Helmholtz Institute Mainz, University of Mainz, 55099 Mainz, Germany}

\emailAdd{hippel@kph.uni-mainz.de}
\emailAdd{jaeger@kph.uni-mainz.de}
\emailAdd{thrae@kph.uni-mainz.de}
\emailAdd{wittig@kph.uni-mainz.de}

\abstract{
          Covariantly smeared sources are commonly used in lattice QCD to
          enhance the projection onto the ground state. Here we investigate the
          dependence of their shape on the gauge field background and find
          that the presence of localized concentrations of magnetic field
          can lead to strong distortions which reduce the
          smearing radii achievable by iterative smearing prescriptions.
          In particular, as $a\to 0$, iterative procedures like Jacobi
          smearing require increasingly large iteration counts in order
          to reach physically-sized smearing radii $r_{\rm sm}\sim 0.5$~fm,
          and the resulting sources are strongly distorted.          
          To bypass this issue, we propose a covariant smearing procedure
          (``free-form smearing'') that allows us to create arbitrarily shaped
          sources, including in particular Gaussians of arbitrary radius.
         }


\preprint{MITP/13-028, HIM-2013-02}

\begin{document}
\maketitle


\section{Introduction}

The construction of interpolating operators optimizing the overlap with the
desired states in correlation functions is a crucial problem in any effort
to extract reliable information from lattice QCD, since contaminations from
excited states can constitute a significant source of systematic error.
This is particularly important in the baryonic sector, where the
signal-to-noise ratio decays rapidly with Euclidean time
(cf.~e.g.~\cite{Capitani:2010sg,Dinter:2011sg,Alexandrou:2013xon,
                Green:2011fg,Green:2012ud,Capitani:2012gj}).
Procedures to isolate the ground and excited states, such as the variational
method
\cite{Michael:1983gevp,Luscher:1990ck,Blossier:2009kd},
also rely on the ability to construct bases of linearly independent operators.
One of the most widely used methods to create operators with improved
projection properties is the use of covariant smearing operations such as
Gaussian or Jacobi smearing
\cite{Gusken:1989ad,Alexandrou:1990dq,Allton:1993wc}.
The intuition behind this is that a hadron should be best described by a
state created by a spatially extended operator rather than a pointlike one,
guided by the principle that the spatial profile of the extended operator
resembles the shape of the hadron in question.
Although smearing techniques can be combined with information on the expected
symmetry properties of hadronic wave functions
\cite{Basak:2005aq,Basak:2005ir},
the procedure remains largely heuristic.

Surprisingly enough, the question of which precise shapes these covariantly
smeared operators have, does not appear to have been studied in any detail.
While it is readily apparent that the application of Gaussian smearing
results in a Gaussian shape on a free gauge configuration, the same does
not necessarily apply in the presence of gauge fields.

In this paper, we study the general properties of covariant smearing and report
on some unexpected discoveries. In section~\ref{sec.smearpaths}, we set up a
general formulation of covariant smearing operations, in order to discuss and
investigate the dependence of the smeared source on the gauge field background.
We find that the existence of localized fluctuations of the chromomagnetic field
can distort the shape of the source away from what one would expect from the
free-field case. In section~\ref{sec.lumps} we study the effects of such
concentrations of field strength, and find them significant, in particular for
high iteration counts of Gaussian smearing, where the resulting sources may,
in fact, bear no resemblance to a Gaussian at all. In particular, this is an
at small lattice spacings, where we find that high iteration counts are needed
to achieve reasonable smearing radii.
To bypass these limitations, in section~\ref{sec.anysmearing}
we introduce a procedure (``free-form smearing'')
to create covariantly smeared sources of arbitrary shapes,
including Gaussians of arbitrary width. We also show
that these sources have much improved projection properties in comparison to
their field-distorted counterparts. Some remaining issues and open questions
are discussed in section~\ref{sec.discussion}.


\section{Covariant smearing: a general framework}\label{sec.smearpaths}

We begin by recapitulating some basic facts about covariant smearing
operations in order to set up a general framework in which the shape of
covariantly smeared sources can be analysed.

A general covariant smearing operation can be written as
\begin{equation}
\widetilde\psi(x) = \sum_y K(x,y) \psi(y)
\end{equation}
with a smearing kernel
\begin{equation}
K(x,y) = \sum_{\mathcal{P}\in P(x,y)} \omega_{\mathcal{P}} U_{\mathcal{P}}
\end{equation}
built from paths $\mathcal{P}$ taken from some set $P(x,y)$ of paths
connecting $x$ and $y$, where this set might typically be chosen so as to
include all paths below a certain length.
Representing a path by a sequence of steps along the coordinate axes,
$\mathcal{P}=(\mu_{\mathcal{P},1},\ldots,\mu_{\mathcal{P},l_\mathcal{P}})$, $1\le|\mu_{\mathcal{P},j}|\le 3$,
and using the conventions that $\hat{e}_{-\mu}=-\hat{e}_\mu$ and
$U_{-\mu}(x)=U_\mu^\dag(x-\hat{e}_\mu)$, we denote by
\begin{equation}
U_{\mathcal{P}} \equiv U_{\mu_{\mathcal{P},1}}(x)\cdots U_{\mu_{\mathcal{P},l_\mathcal{P}}}(x+\sum_{j=1}^{l_\mathcal{P}-1} \hat{e}_{\mu_{\mathcal{P},j}})
\end{equation}
the product of links required to parallel transport a spinor along that
path. 

For the usual case of a point source, $\psi(y)=\phi_0\delta(y-y_0)$ with
a normalized SU(3) vector $\phi_0\in\Cset^3$, $\norm{\phi_0}=1$, we can
perform the sum over $y$, in which case the explicit form of the
covariant smearing operation reads
\begin{equation}
\widetilde\psi(x) = \sum_{\mathcal{P}\in P(x,y_0)} \omega_{\mathcal{P}}
U_{\mathcal{P}}\phi_0\,.
\label{eqn:smear:gen}
\end{equation}

In practice, covariant smearing operations are not implemented as sums over
paths, since the number of paths to take into account would generally be
unmanageable (e.g.\ for smearing a point source over a $5^3$ volume,
i.e.\ a very small smearing radius of $r\le 2a$, there are $14,005$
distinct paths of length up to $6$ that can contribute).
Instead, iterative procedures like Jacobi or Gaussian smearing
\cite{Gusken:1989ad,Alexandrou:1990dq}
are generally employed, where the smeared field is derived from the
unsmeared one by the repeated application of a derivative or hopping operator
$H$ in the form
\begin{equation}
\widetilde\psi = C \left(1+\kappa_{\rm G} H\right)^n \psi \,.
\label{eqn:smear:gauss}
\end{equation}
However, in theory, it is relatively simple to perform the expansion
of eqn.~(\ref{eqn:smear:gauss}) to obtain the form eqn.~(\ref{eqn:smear:gen}),
for example, by using a similar bottom-up algorithm as has been applied to the
perturbative expansion of lattice QCD actions
\cite{Hart:2009nr};
in practice, the number of paths grows exponentially with $n$, so that
the enumeration of all paths contributing for practically relevant values
of $n$ becomes infeasible.

A useful measure of the spatial extent of a smeared source is given by the
smearing radius defined through
\begin{equation}
r_{\rm sm}^2 = \frac{\displaystyle\sum_{x} |x-y_0|^2 \norm{\tilde\psi(x)}^2}%
                    {\displaystyle\sum_{x} \norm{\tilde\psi(x)}^2}\,,
\end{equation}
where $y_0$ is the position of the source. The norm $\lVert\tilde\psi(x)\rVert$
of the smeared source at a given spatial point $x$ can thus be seen as
determining the weight of the point $x$, and the norm can thus be identified
with the general ``shape'' of the source.

Taking the norm of the smeared source in eqn.~(\ref{eqn:smear:gen}), we find
\begin{eqnarray}
\norm{\widetilde\psi(x)}^2 &=& \norm{\sum_{\mathcal{P}\in P(x,y_0)} \omega_{\mathcal{P}} U_{\mathcal{P}}\phi_0}^2 \\
&=& \sum_{\mathcal{P}\in P(x,y_0)} \omega_{\mathcal{P}}^2 + \sum_{\mathcal{P}_i\not=\mathcal{P}_j\in P(x,0)} \omega_{\mathcal{P}_i}\omega_{\mathcal{P}_j} \phi_0^\dag U_{\mathcal{P}_i}^{\dag}U_{\mathcal{P}_j}\phi_0\,.
\end{eqnarray}
Since the smearing does not affect the Dirac structure of the source, a full
set of sources can be obtained by averaging over the colour components, giving
\begin{equation}
\norm{\widetilde\psi(x)}^2 = \sum_{\mathcal{P}_i\in P(x,y_0)} \omega_i^2 + 2 \sum_{\mathcal{P}_i,\mathcal{P}_j\in P(x,y_0),\atop i<j} \omega_{\mathcal{P}_i}\omega_{\mathcal{P}_j} \textrm{Re tr}\left[U_{\mathcal{P}_i}^{\dag}U_{\mathcal{P}_j}\right]\,.
\end{equation}
The presence of the last term implies that the shape of the source created
by the application of the smearing operation will depend on the gauge field
background through
\begin{equation}
\textrm{Re tr}\left[U_{\mathcal{P}_i}^{\dag}U_{\mathcal{P}_j}\right]=\int_{\Omega_{ij}}\textrm{Re tr}\left[B^2+\ldots\right] \,,
\label{eqn:smear:ab}
\end{equation}
where $\Omega_{ij}$ is the surface bounded by ${P}_i$ and ${P}_j$.
Therefore, we expect that localized regions of strong chromomagnetic fields
will affect the shape and, in turn, also the projection properties of a
covariantly smeared source.


\section{Empirical evidence for magnetic distortion}\label{sec.lumps}

\begin{figure}
\begin{center}
\includegraphics[width=0.49\textwidth,keepaspectratio=]{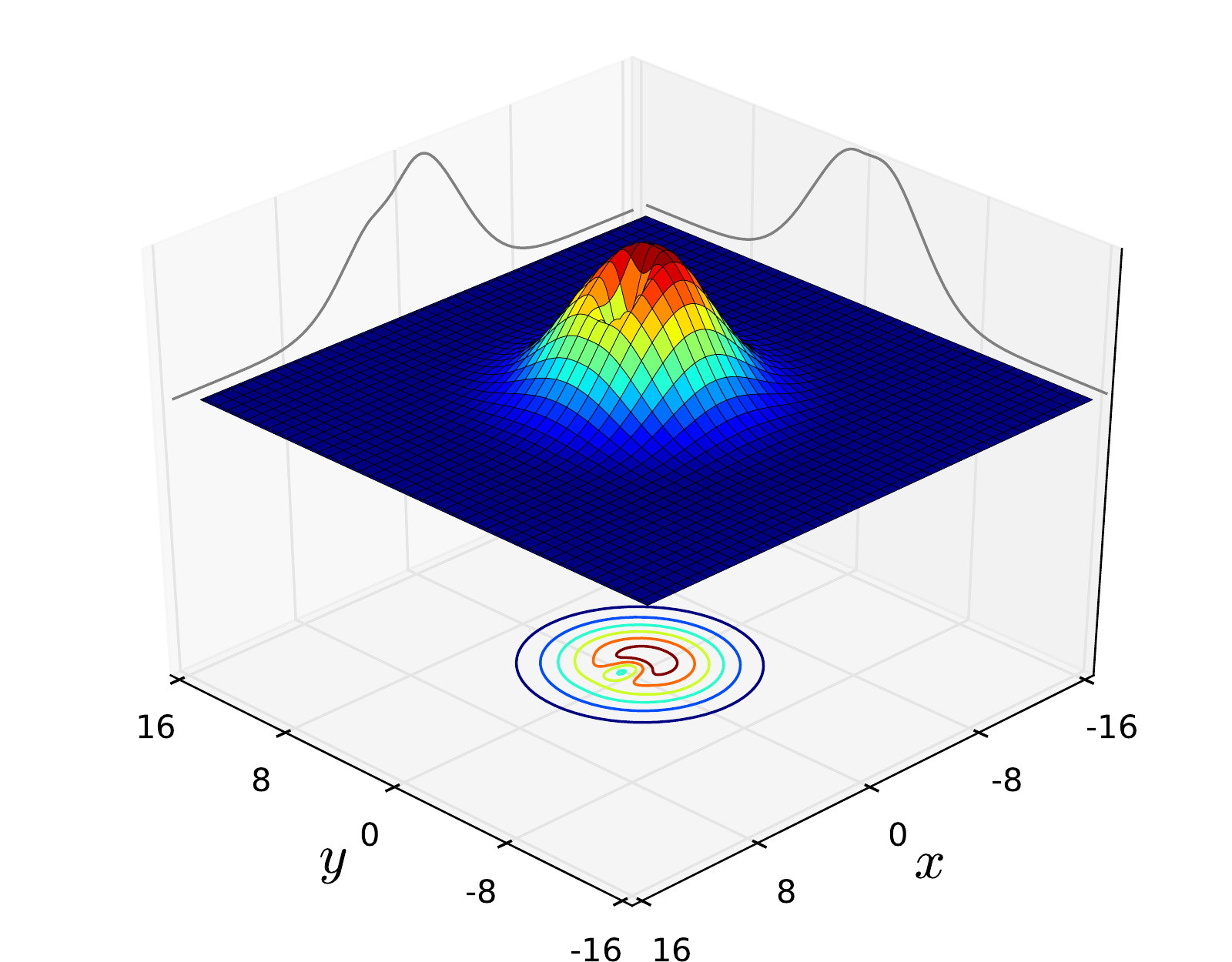}
\includegraphics[width=0.49\textwidth,keepaspectratio=]{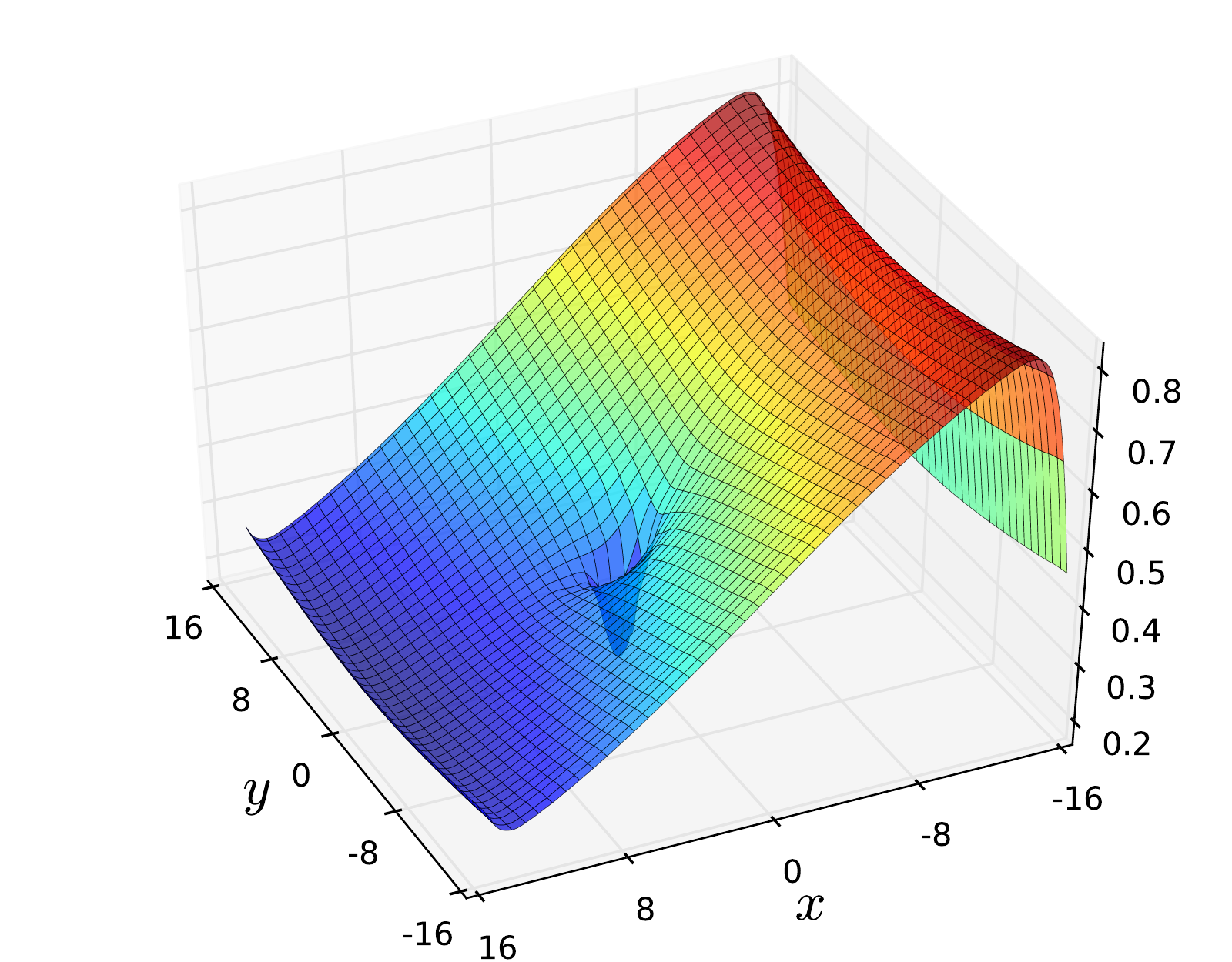}
\end{center}
\caption{{\em left:} the shape of a Gaussian source that has been distorted
                     by an artificially introduced magnetic flux
                     on an otherwise free (unit) configuration;
         {\em right:} the ratio between the Gaussian source distorted by the
                      flux
                      and the same Gaussian source on a free (unit)
                      configuration; it can be seen that besides creating a
                      strong local suppression, the presence of the
                      flux
                      shifts and distorts the source globally.}
\label{fig:singlelink}
\end{figure}

To demonstrate that localized magnetic flux distorts source shapes, we consider
first the case in which a highly localized concentration of magnetic field
strength has been inserted into an otherwise trivial background by hand.
Specifically, we consider the gauge configuration
\begin{equation}
U_\mu(x) = \left\{\begin{array}{l} u_0,~\mu=3,~x=(0,0,0,0), \\
                                   \openone,~\text{else,}\end{array}\right.
\end{equation}
where $u_0$ is chosen far from the identity, and apply Gaussian smearing to
the source $\psi(x)=\phi_0\delta(x-y_0)$, where $y_0=(0,L/2,L/2,L/2)$ and
$\phi_0$ is a unit vector in Dirac-colour space. For this configuration,
the change of the norm of the smeared source compared to the free case
is largely a function of the number of contributing paths that pass through
the non-unit link.

\begin{figure}
\begin{center}
\includegraphics[width=0.49\textwidth,keepaspectratio=]{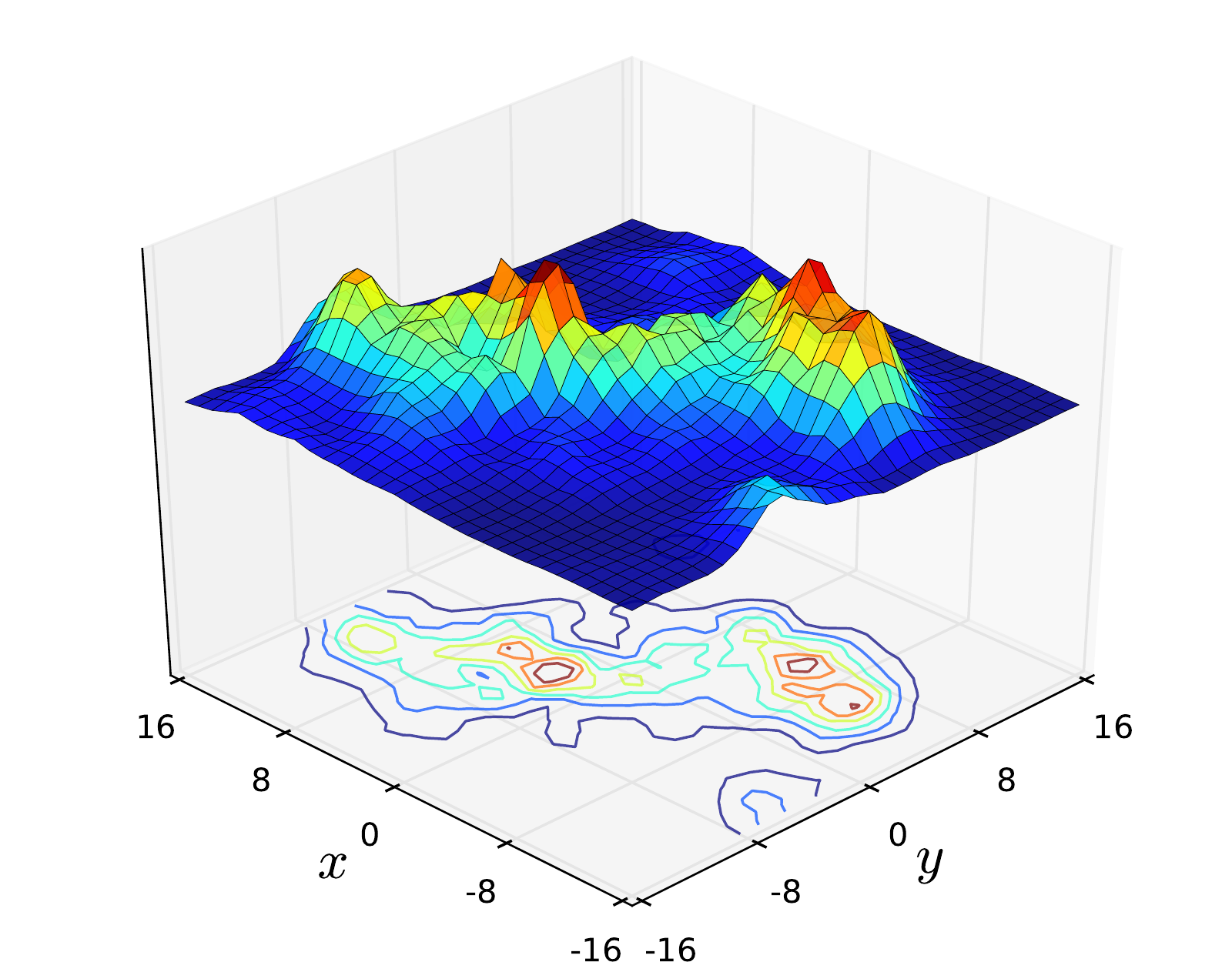}
\includegraphics[width=0.49\textwidth,keepaspectratio=]{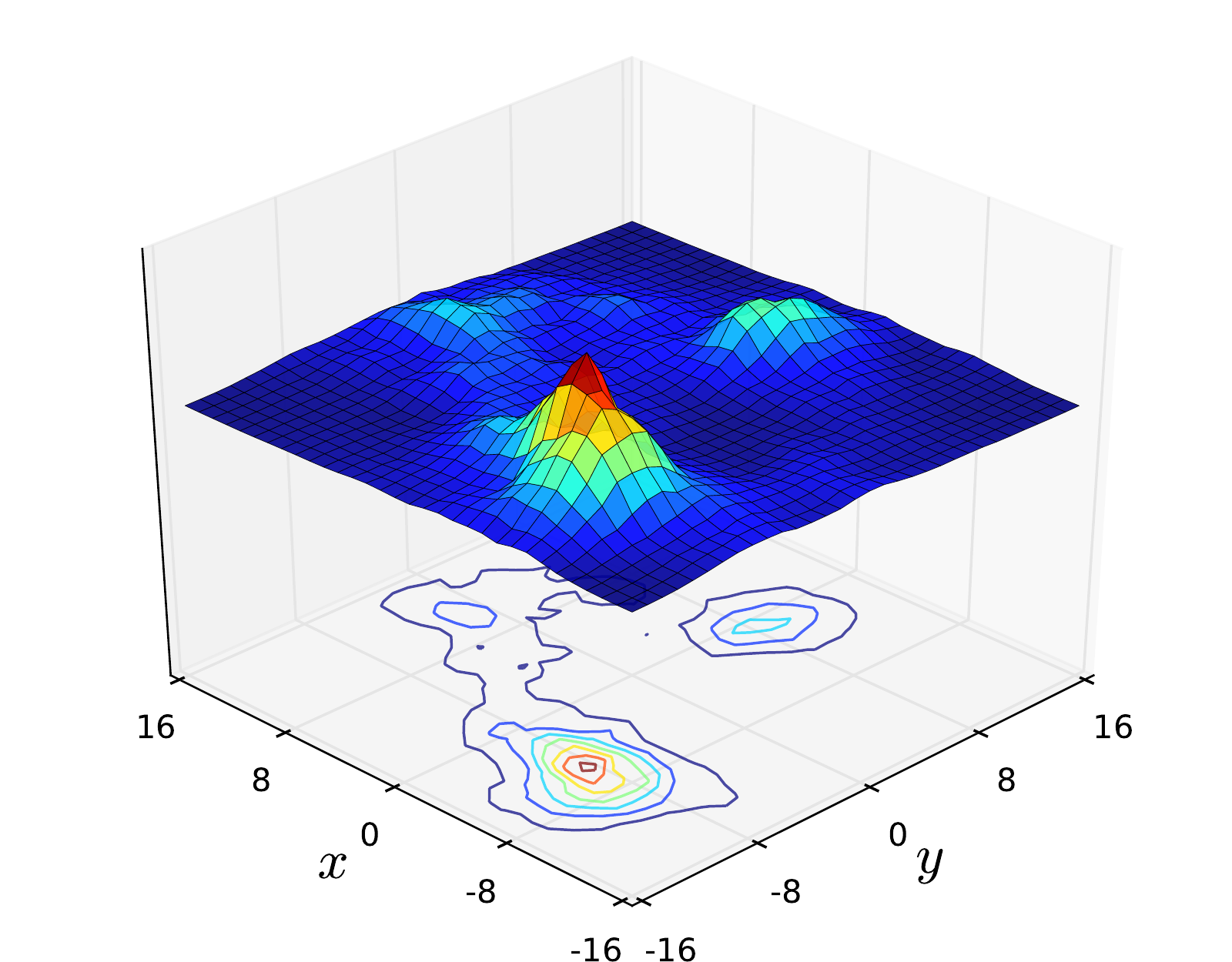}
\end{center}
\caption{Examples of strongly distorted source shapes encountered by
         applying eqn.~(\ref{eqn:smear:gauss}) with $\kappa_{\rm G}=2.9$, $n=640$
         to configurations from an $a=0.063$~fm ensemble
         with $N_{\rm f}=2$ flavours of O($a$)-improved Wilson fermions.
         The corresponding smeared source shape in the free theory
         would be a broad Gaussian centered on the middle point of this plot.}
\label{fig:realheffas}
\end{figure}

The resulting source shape can be seen in figure~\ref{fig:singlelink}.
Two features that can be clearly observed are the strong depression of the
source at the ends of the non-unit link, as well as the global distortion
of the whole source.
Similar distorting effects will also occur in configurations forming part of
actual lattice QCD ensembles, as is demonstrated by
figure~\ref{fig:realheffas}. We see that for large iteration numbers,
the distorting effects can be so strong that the resulting shape is very
decidedly non-Gaussian.

\begin{figure}
\begin{center}
\includegraphics[width=0.49\textwidth,keepaspectratio=]{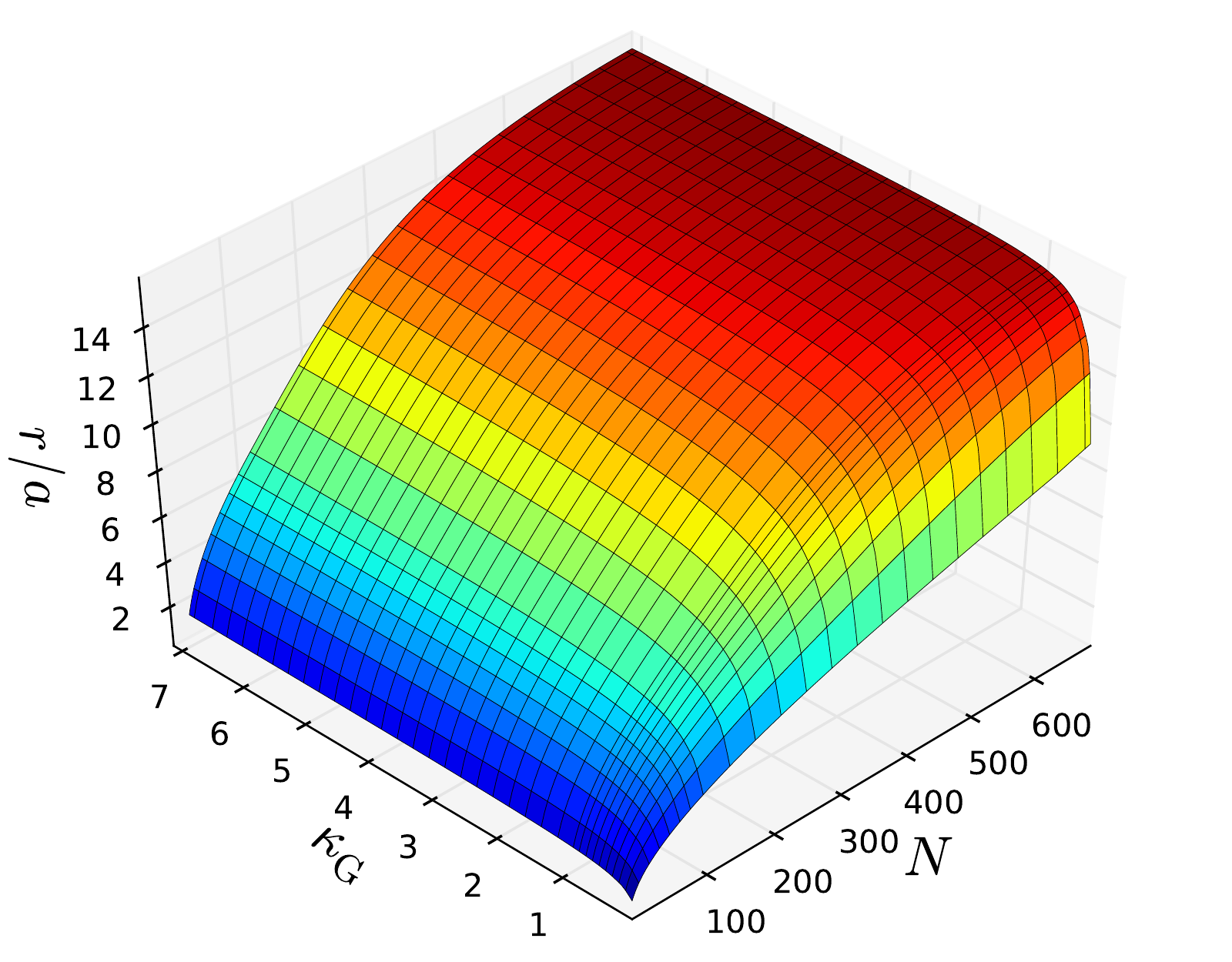}
\includegraphics[width=0.49\textwidth,keepaspectratio=]{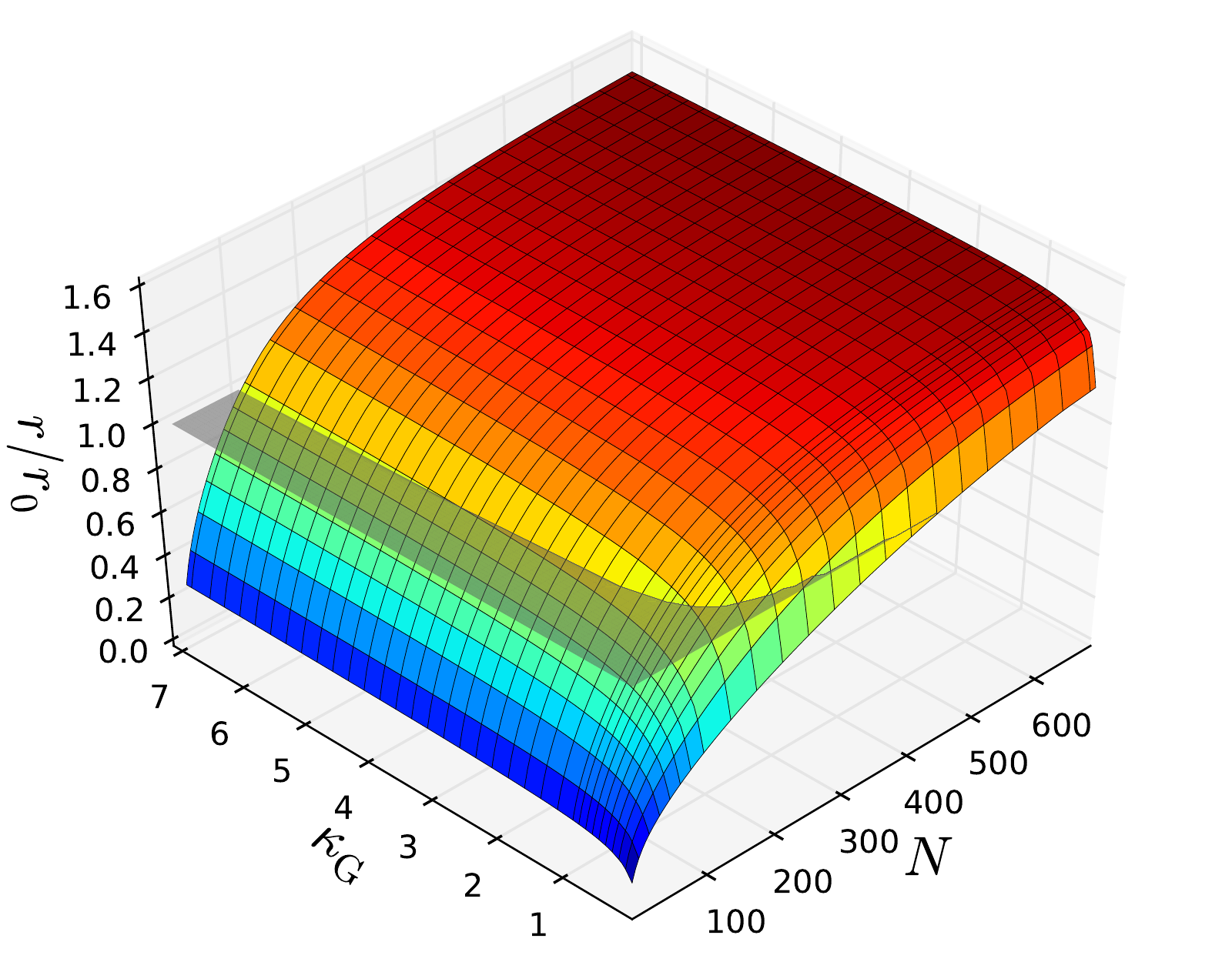}\\
\includegraphics[width=0.49\textwidth,keepaspectratio=]{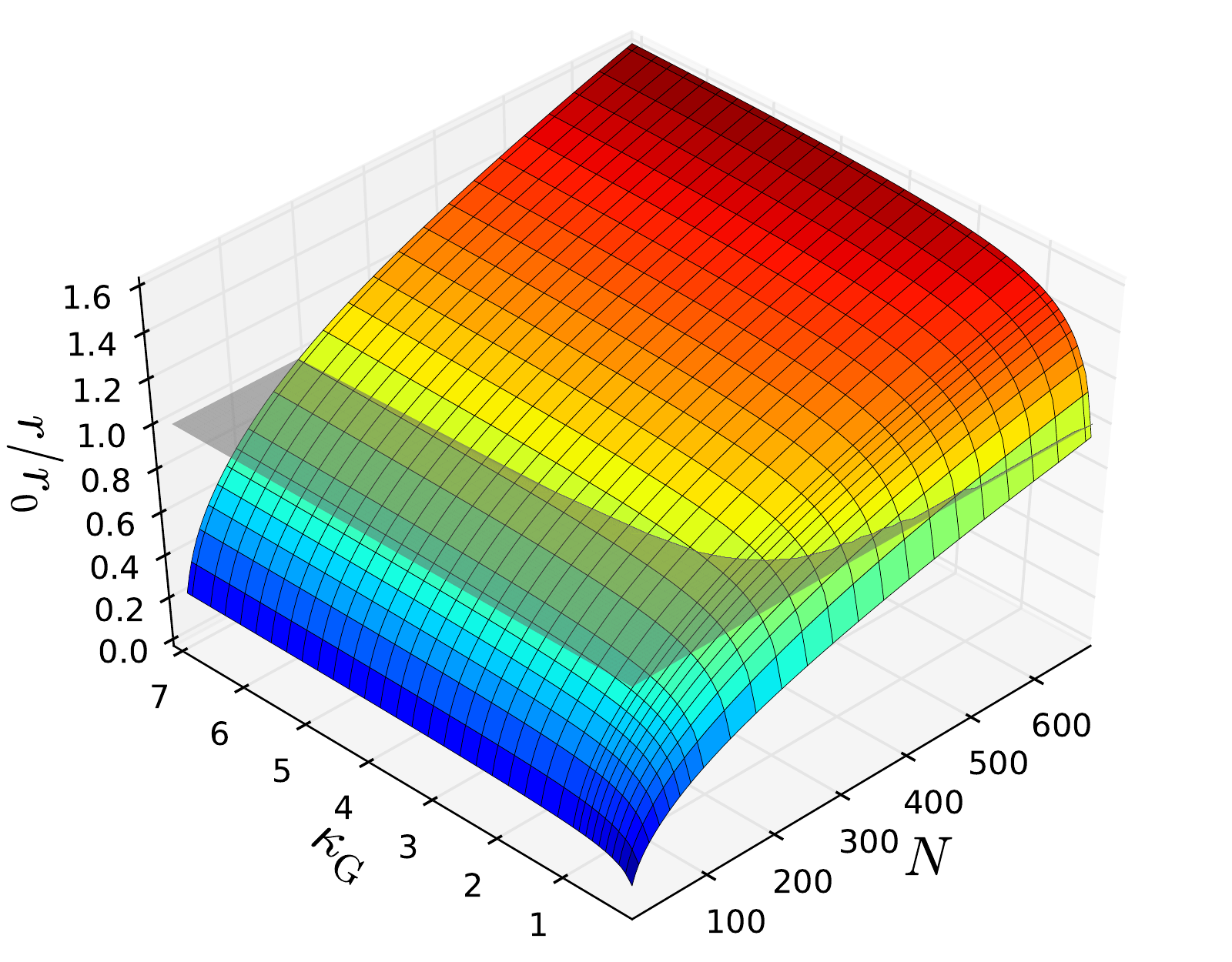}
\includegraphics[width=0.49\textwidth,keepaspectratio=]{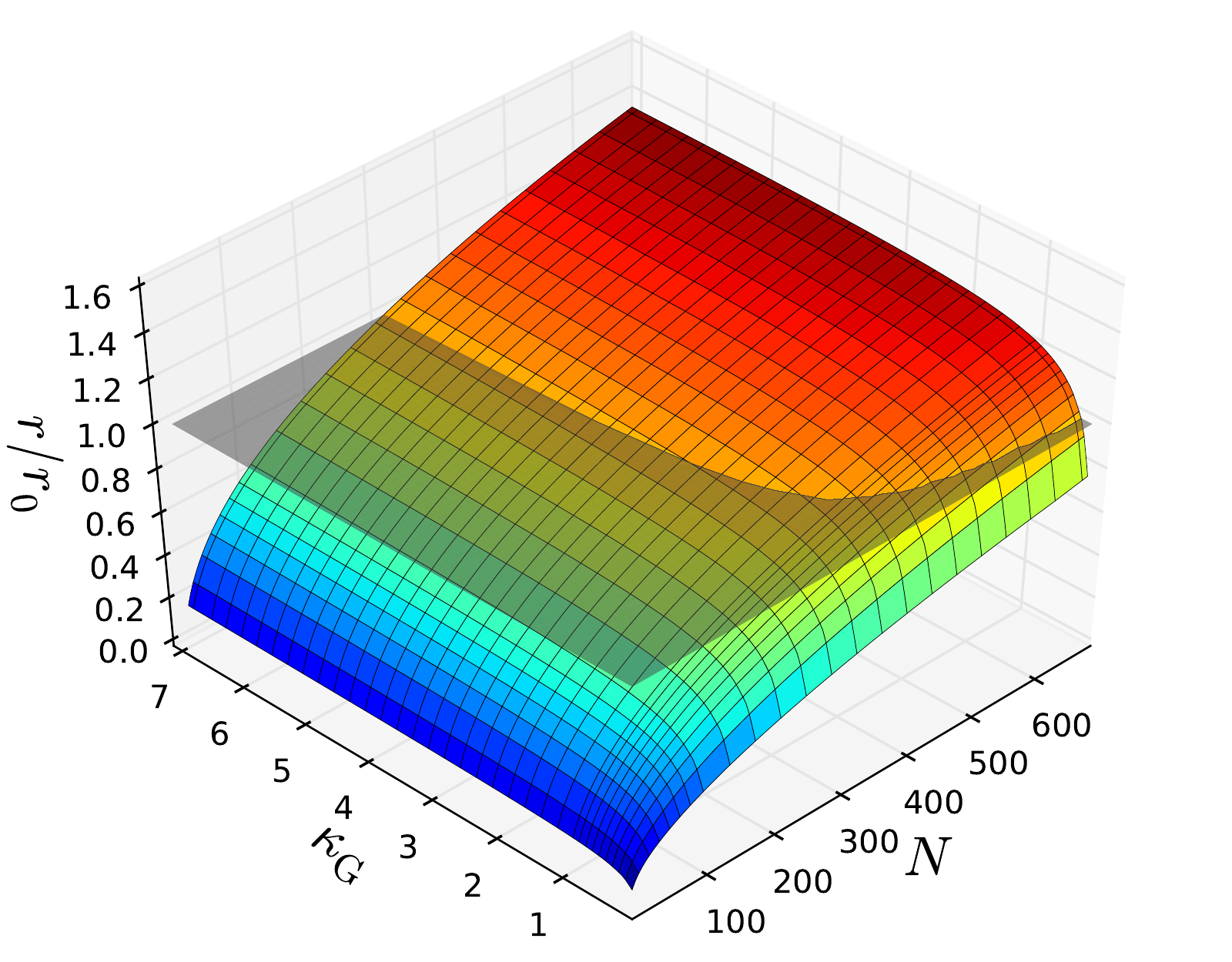}
\end{center}
\caption{Gaussian ``smearscapes'' for the free case (top left),
         an ensemble with $a=0.079$~fm (top right),
         an ensemble with $a=0.063$~fm (bottom left), and
         an ensemble with $a=0.050$~fm (bottom right).
         Shown is the smearing radius $r_{\rm sm}$ as a function
         of both $\kappa_{\rm G}$ and the iteration number $n$.
         The semitransparent horizontal plane indicates where
         $r_{\rm sm}=0.5$~fm is achieved in the interacting case.
         The increasing suppression of the radius as $a\to 0$ can
         be clearly seen. For the latter three cases, the smearing
         radius is plotted in units of $r_0$.}
\label{fig:smearscape}
\end{figure}

\begin{figure}
\begin{center}
\includegraphics[width=0.49\textwidth,keepaspectratio=]{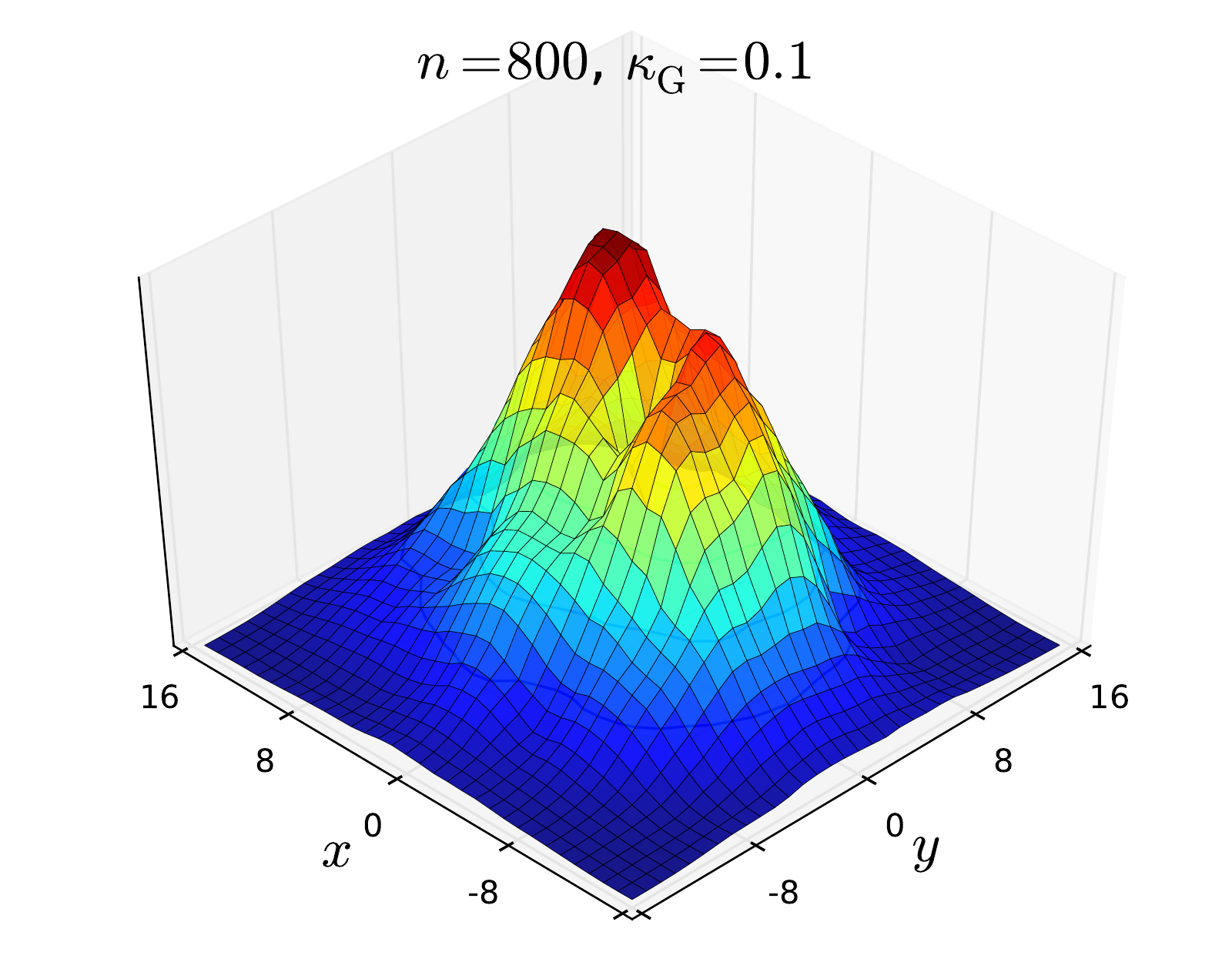}
\includegraphics[width=0.49\textwidth,keepaspectratio=]{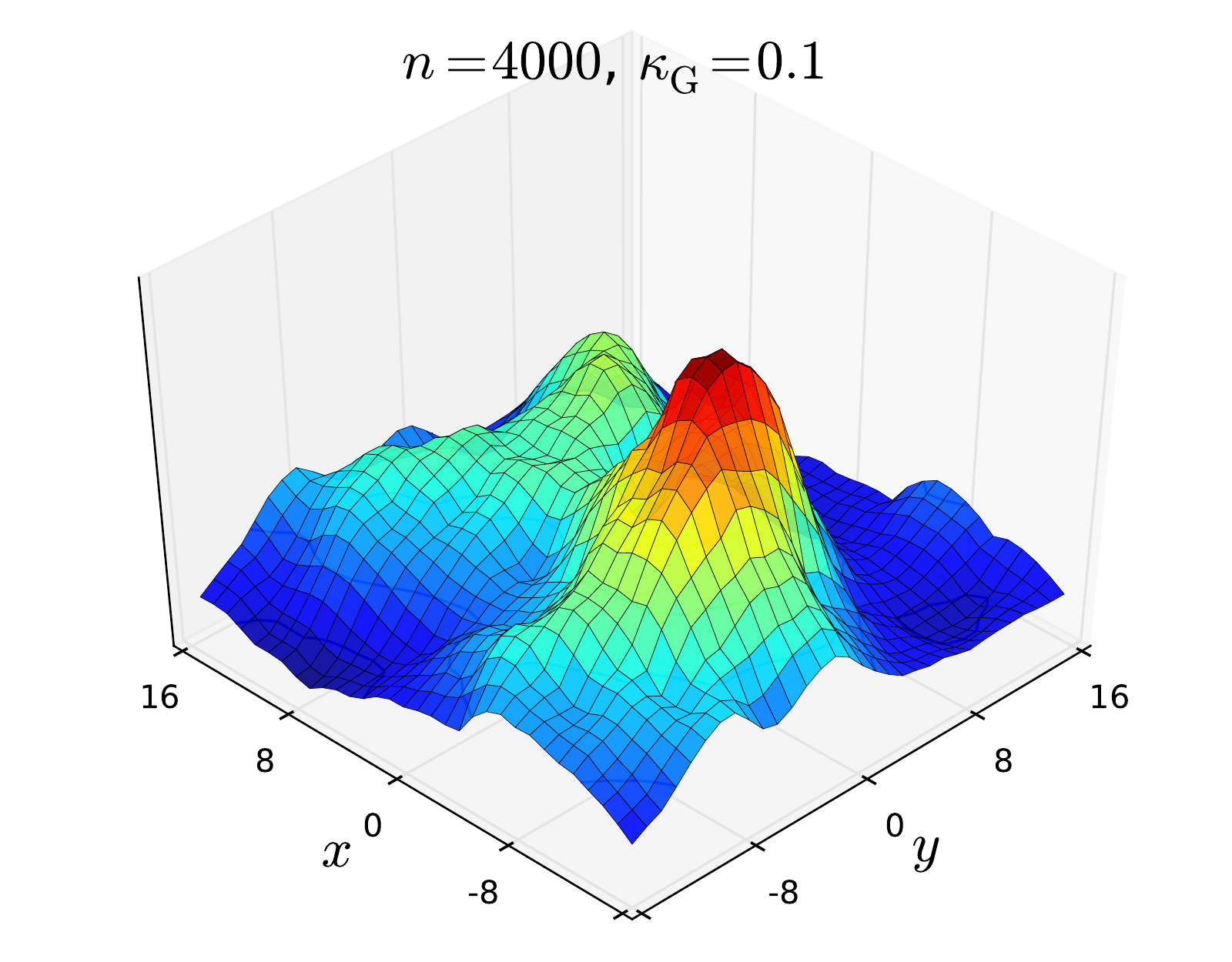}\\
\includegraphics[width=0.49\textwidth,keepaspectratio=]{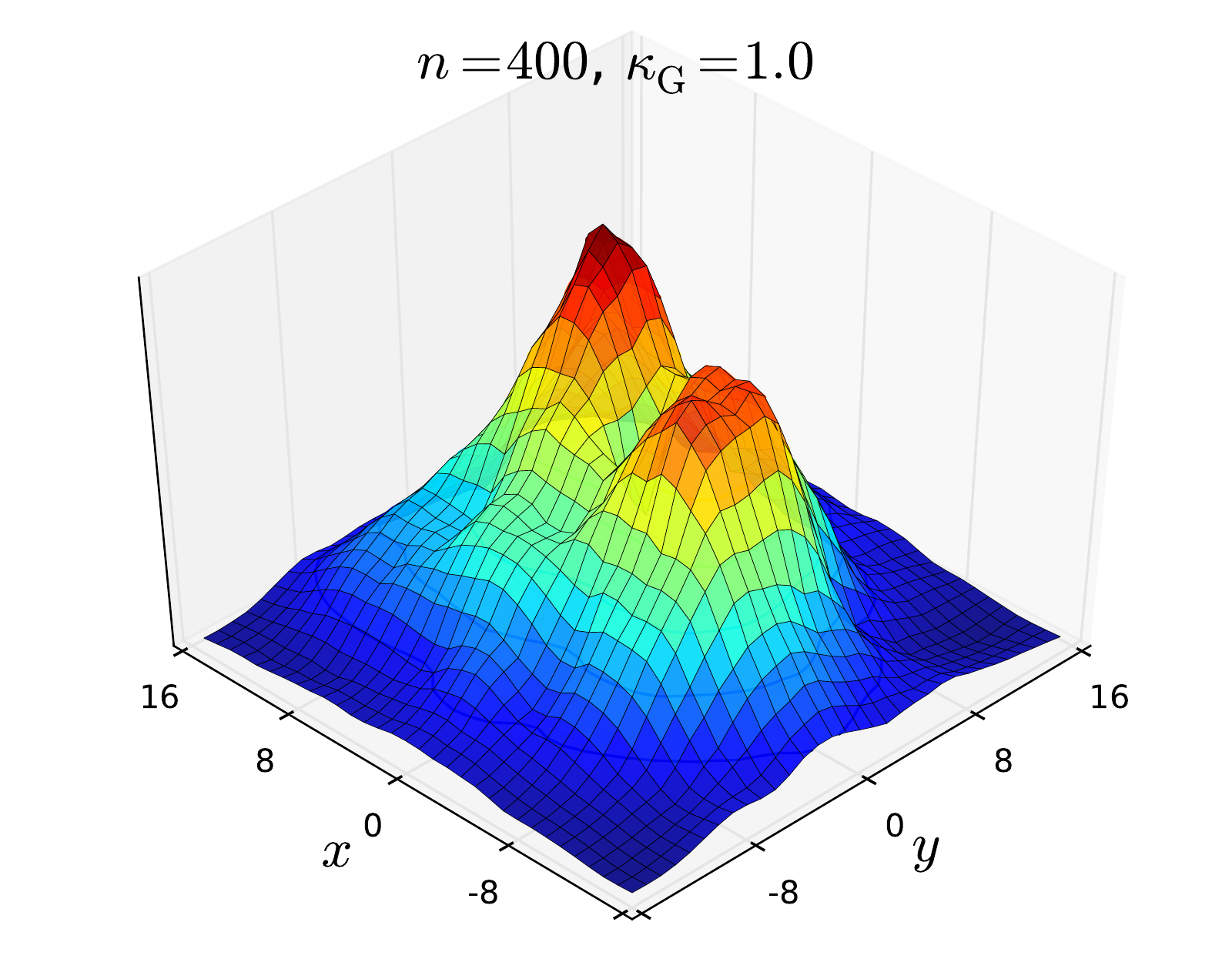}
\includegraphics[width=0.49\textwidth,keepaspectratio=]{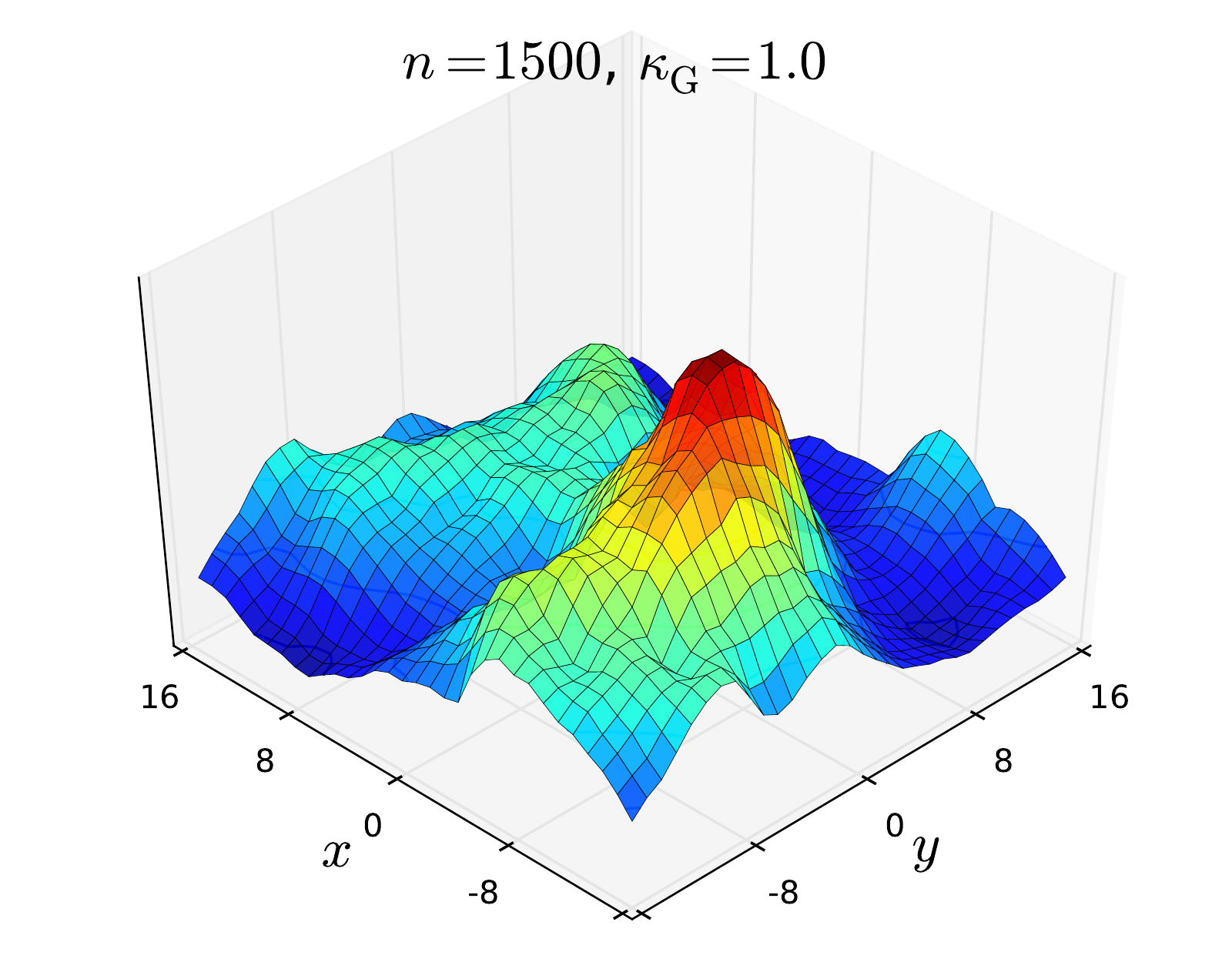}\\
\includegraphics[width=0.49\textwidth,keepaspectratio=]{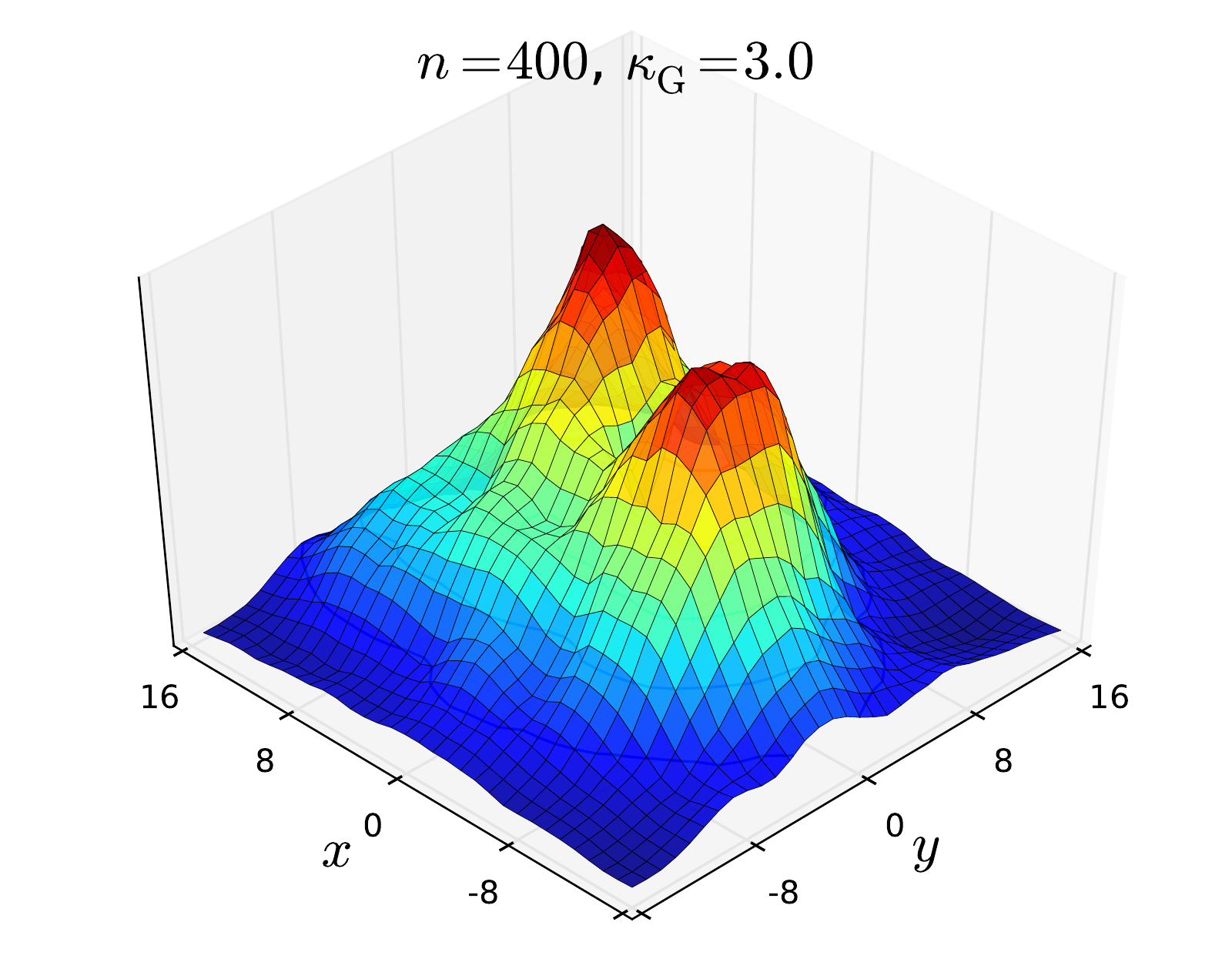}
\includegraphics[width=0.49\textwidth,keepaspectratio=]{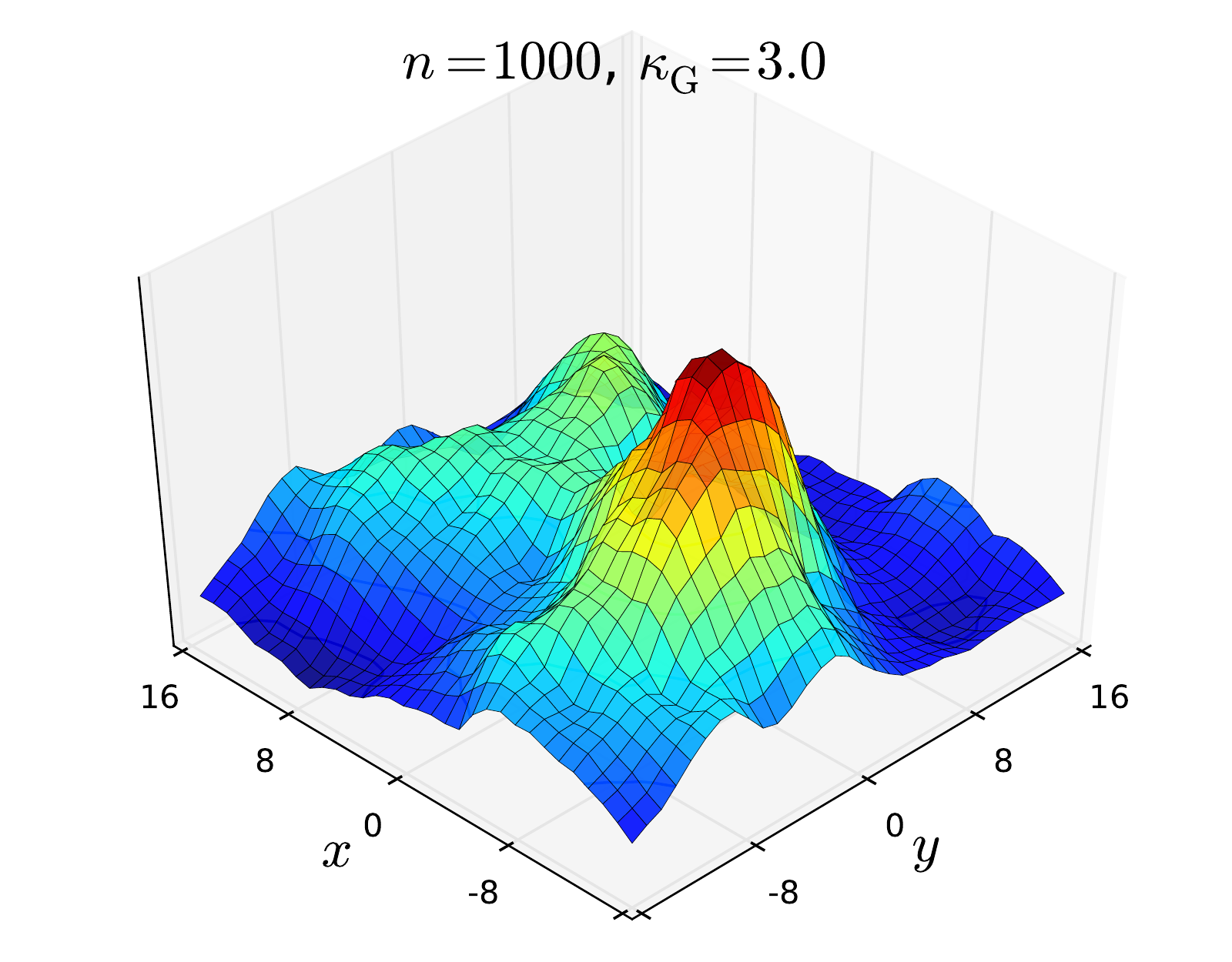}
\end{center}
\caption{Shapes of covariantly smeared sources on the same
         $a=0.063$~fm gauge configuration
         for three different parameter sets
         (from top to bottom: $\kappa_{\rm G}=0.1$, $1.0$, $3.0$).
         On the left, a moderate number of iterations is used, whereas on the
         right, a large number of iterations has been applied, with the
         iteration numbers at each value of $\kappa_{\rm G}$ chosen so as to
         correspond to approximately the same two smearing radii.
         It can be seen rather clearly that with increased smearing the source
         is attracted away from the centre by the influence of the gauge field
         background, and that for similar smearing radii, similar shapes are
         observed, regardless of the precise parameters used.
         }
\label{fig:parmdep}
\end{figure}

The presence of these distortions creates problems when attempting to create
Gaussian sources with large radii by the iteration of
eqn.~(\ref{eqn:smear:gauss}), since at large iteration numbers the number of
paths contained in the smearing operation grows rapidly, thereby increasing
the global influence of the local magnetic flux
as more and more paths routed through them start contributing
to the source at all points. A consequence of this is that 
the radii found on actual gauge
ensembles will be significantly smaller and grow much more slowly with the
iteration number than the corresponding radii in the free theory.
This problem can be clearly seen in the ``smearscapes'' shown in
figure~\ref{fig:smearscape}; for the actual gauge ensembles, reaching source
radii larger than $r_{\rm sm}~\ltsim~0.5$~fm, becomes essentially impossible
-- even if one were willing to pay the required price in terms of iterations,
the resulting sources would have surreal shapes (cf.
figure~\ref{fig:realheffas}) bearing absolutely no similarity
to a Gaussian on most configurations.

Further evidence that the limits on the achievable smearing radii come from
the distortion of the source shapes comes from studying the dependence of the
source shape on the parameters $\kappa_{\rm G}$, $n$ of the Gaussian smearing.
In figure~\ref{fig:parmdep}, moderately and highly smeared sources for three
parameter sets corresponding to approximately the same pair of radii are shown.
It can be seen that the same distortion occurs in all three cases.


\section{Covariantly smearing with arbitrary shapes}\label{sec.anysmearing}

In order to overcome some of the limitations that these effects may impose
on achievable smearing radii, as well as to avoid the need to gauge-fix
an ensemble for introducing non-Gaussian source shapes such as
hydrogenic wavefunctions, let us return to eqn.~(\ref{eqn:smear:gen})
and consider the possibility of creating an arbitrarily shaped source
satisfying $\lVert\tilde\psi(x)\rVert = |f(x-y_0)| \norm{\phi_0}$ with
an arbitrary function $f$. In a gauge-fixed framework, we might simply use
$\tilde\psi(x) = f(x-y_0) \phi_0$ to this end, which would correspond to
setting $K(x,y)=f(x-y)$ in eqn.~(\ref{eqn:smear:gen}). This procedure
can be rendered gauge covariant by singling out one particular path
$\mathcal{P}_{xy}$ from $y$ to $x$ and defining
\begin{equation}
K(x,y) = f(x-y) U_{\mathcal{P}_{xy}} \,.
\end{equation}
This would seem to entirely avoid the influence of magnetic fluxes,
since there are no different paths that could interfere with each other,
and hence the desired relation
$\lVert\tilde\psi(x)\rVert = |f(x-y_0)| \norm{\phi_0}$
holds regardless of the gauge field.

However, this radical solution is not actually practical, since singling out
a specific path makes the source much too vulnerable to ultraviolet fluctuations
of the individual links, leading to large statistical errors in the correlation
functions built from such sources. The same applies if one averages over a
moderately-sized set of paths, for example the one created by reflecting a
single path along the different coordinate axes.

As an alternative solution, we should therefore consider defining a sufficiently
large collection of paths by expanding eqn.~(\ref{eqn:smear:gauss}) with a
relatively modest iteration count (chosen, for example, as the minimum needed
to be able to reach every point on the lattice) into the form of
eqn.~(\ref{eqn:smear:gen}), and merely replacing the weights
$\omega_{\mathcal{P}}$ so as to reshape the source to obey
$\langle\lVert\tilde\psi(x)\rVert\rangle = |f(x-y_0)| \norm{\phi_0}$ in the gauge
average. Unfortunately, this solution is rendered non-viable by the prohibitive
cost of enumerating all paths for the iteration numbers required on typical
lattices. We can, however, bypass the need to actually enumerate all paths by
applying the following procedure:
\begin{itemize}
\item calculate $\psi'(x) = [\left(1+\kappa_{\rm G} H\right)^n]_{x,y}\delta(y-y_0)\phi_0$, with $n$ chosen large enough to touch all points by at least one path,
\item determine $N(x) = \langle\norm{\psi'(x)}\rangle$,
\item define $\tilde\psi(x) = \psi'(x) f(x-y_0)/N(x)$,
\end{itemize}
in order to create an arbitrarily-shaped source.
We shall call this procedure ``free-form smearing''.%
\footnote{An obvious alternative would be to enforce
          $\langle\lVert\tilde\psi(x)\rVert^2\rangle = 
                       |f(x-y_0)|^2 \norm{\phi_0}^2$
          by defining $N(x) = \sqrt{\langle\norm{\psi'(x)}^2\rangle}$.}

In particular, free-form smearing can be used to create Gaussian source shapes of
arbitrary radius without encountering significant distortions of the source
shape, thus bypassing the limitations on the source radius found in the preceding
section. It can also be used to create anisotropic sources similar to
those considered to study moving hadrons in
\cite{DellaMorte:2012xc},
or to build a basis of operators for the application of methods to extract
excited state spectra
\cite{Michael:1983gevp,Luscher:1990ck,Dudek:2007wv,Blossier:2009kd,
      vonHippel:2007ar,Hornbostel:2011hu,Fleming:2009wb}.
In the latter case, the ability of free-form smearing to create shapes other than
Gaussian is particularly useful, since it enables the use of ans\"atze with nodes
, such as  hydrogenic wavefunctions, without requiring us to fix the gauge
(as had to be done for that purpose in, e.g.,
\cite{Dowdall:2011wh}).

\begin{figure}
\begin{center}
\includegraphics[width=0.49\textwidth,keepaspectratio=]{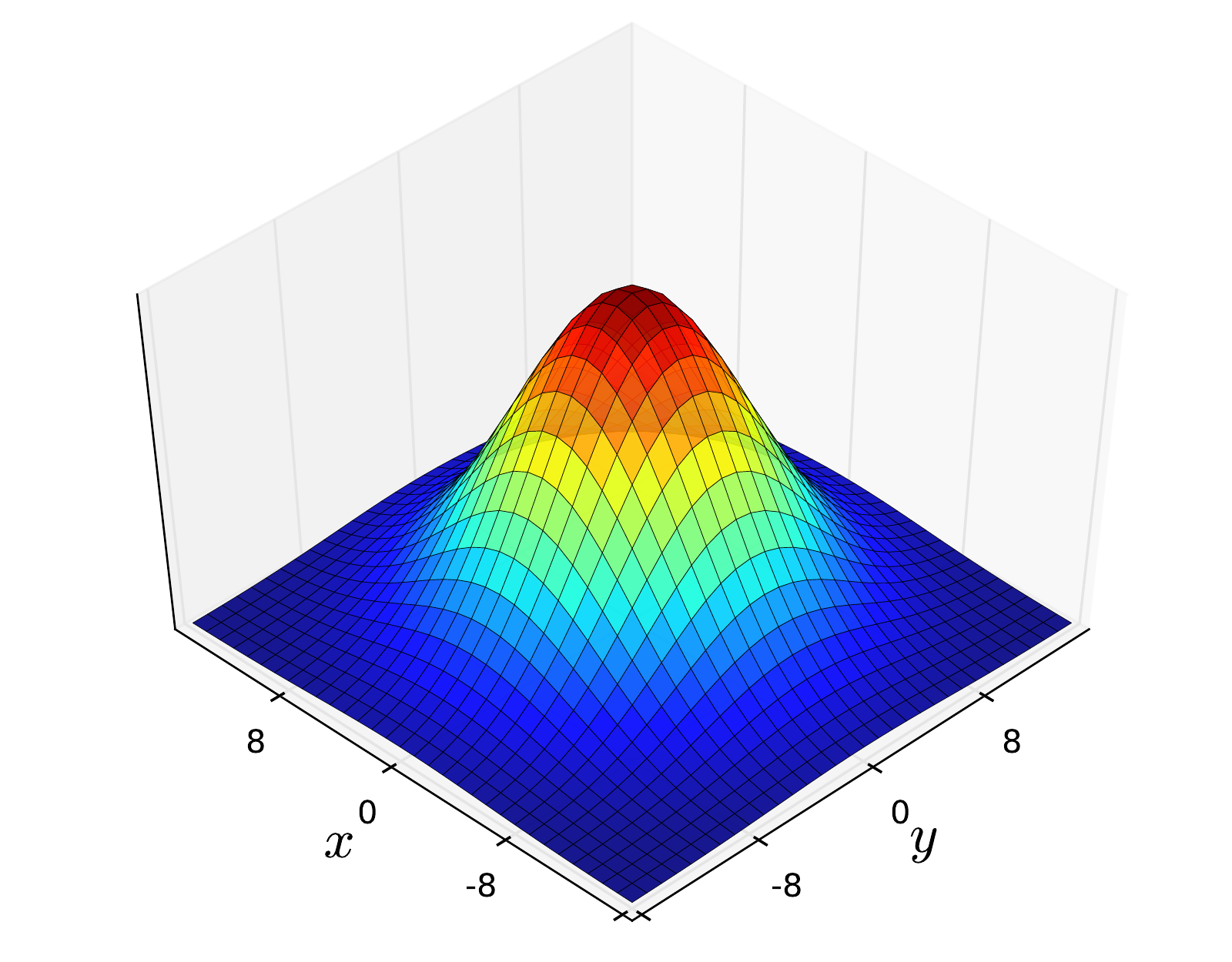}
\includegraphics[width=0.49\textwidth,keepaspectratio=]{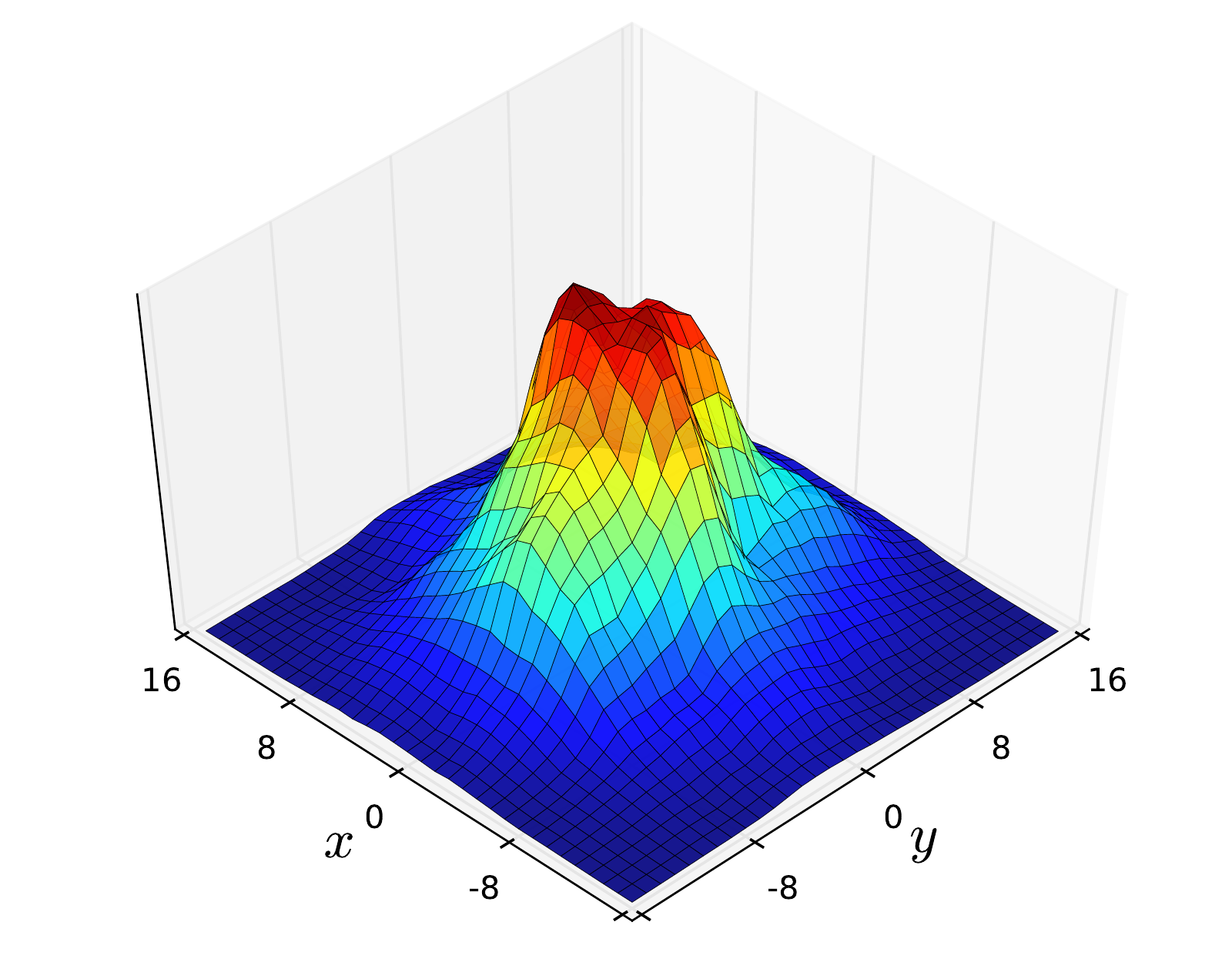}\\
\includegraphics[width=0.49\textwidth,keepaspectratio=]{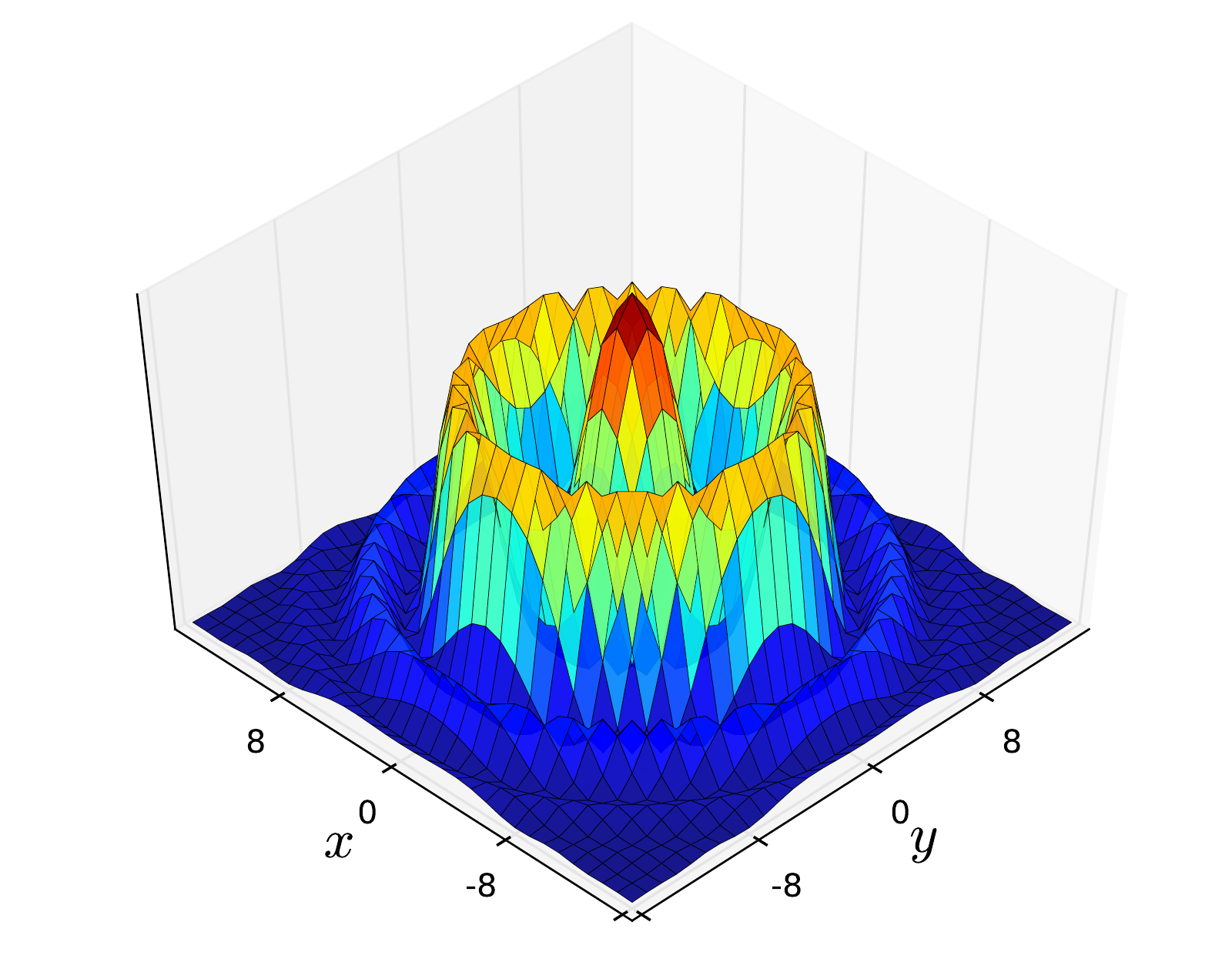}
\includegraphics[width=0.49\textwidth,keepaspectratio=]{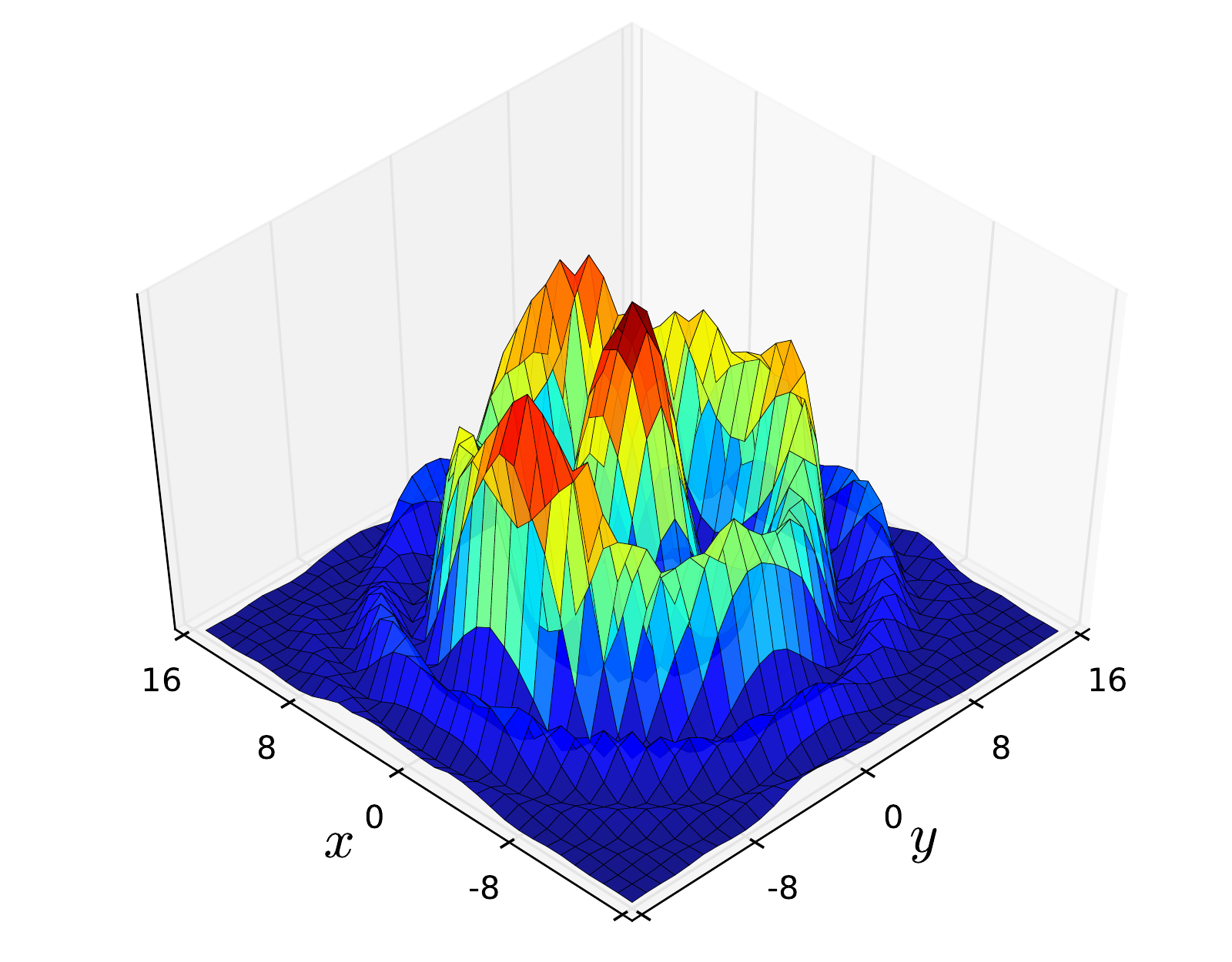}
\end{center}
\caption{A comparison of the shapes of free-form smeared sources on a unit
         configuration (left) and a configuration taken from an $a=0.063$~fm
         ensemble (right); shown are an free-form smeared Gaussian source (top)
         and a source with
         $f(z)=(1 - \sin(\pi r^2/50a^2))\exp(-r^2/72a^2)$
         demonstrating the possibility to free-form smear arbitrary source
         shapes (bottom).}
\label{fig:shapecomp}
\end{figure}

A comparison of the ideal source shape $f(x-y_0)$ and the result of applying
free-form smearing to a given gauge configuration can be seen in
figure~\ref{fig:shapecomp}
for the case of both a Gaussian and an arbitrarily chosen source shape
$f(z)=(1 - \sin(\pi r^2/50a^2))\exp(-r^2/72a^2)$ displaying nodes.
It can be seen that in spite of the fluctuations
due to the gauge fields, the free-form smeared shape is well-reproduced even
on an individual gauge configuration.

\begin{figure}
\begin{center}
\includegraphics[width=0.69\textwidth,keepaspectratio=]{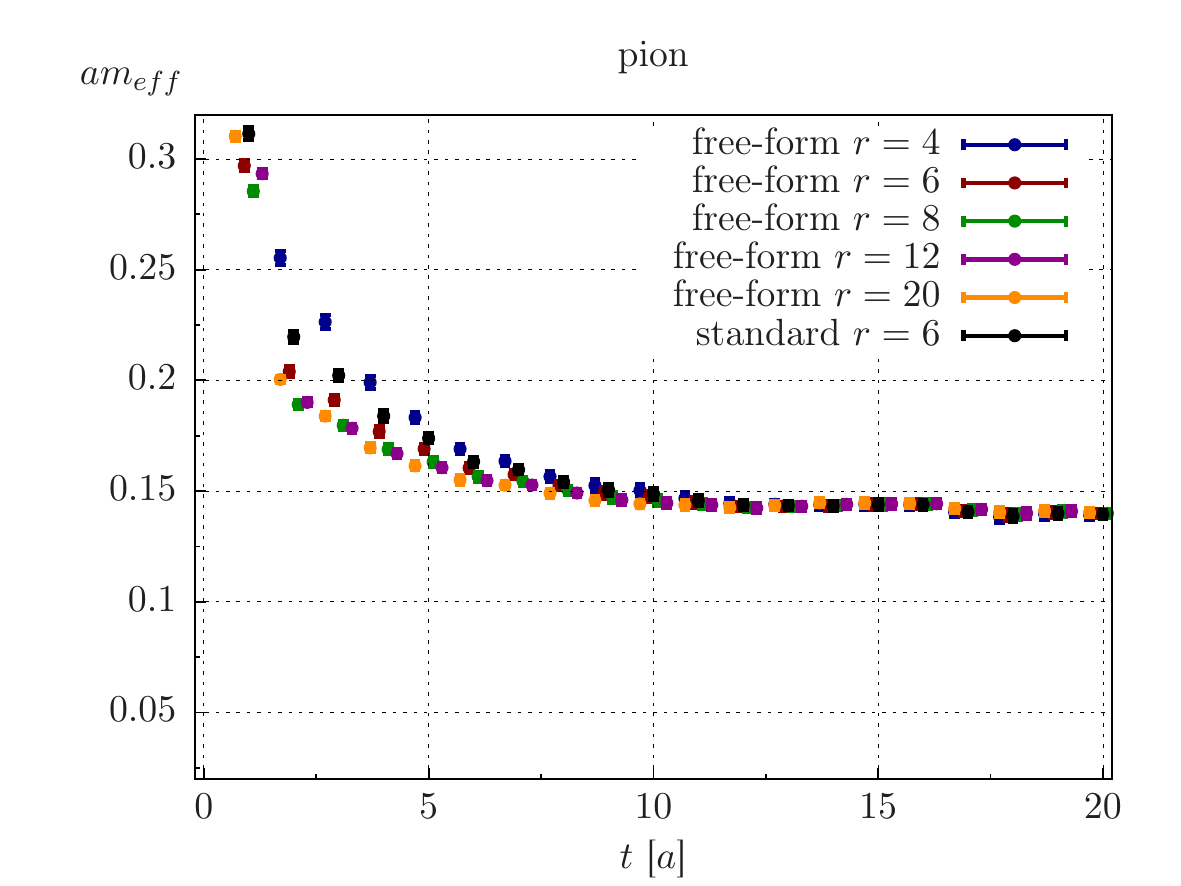}\\
\includegraphics[width=0.69\textwidth,keepaspectratio=]{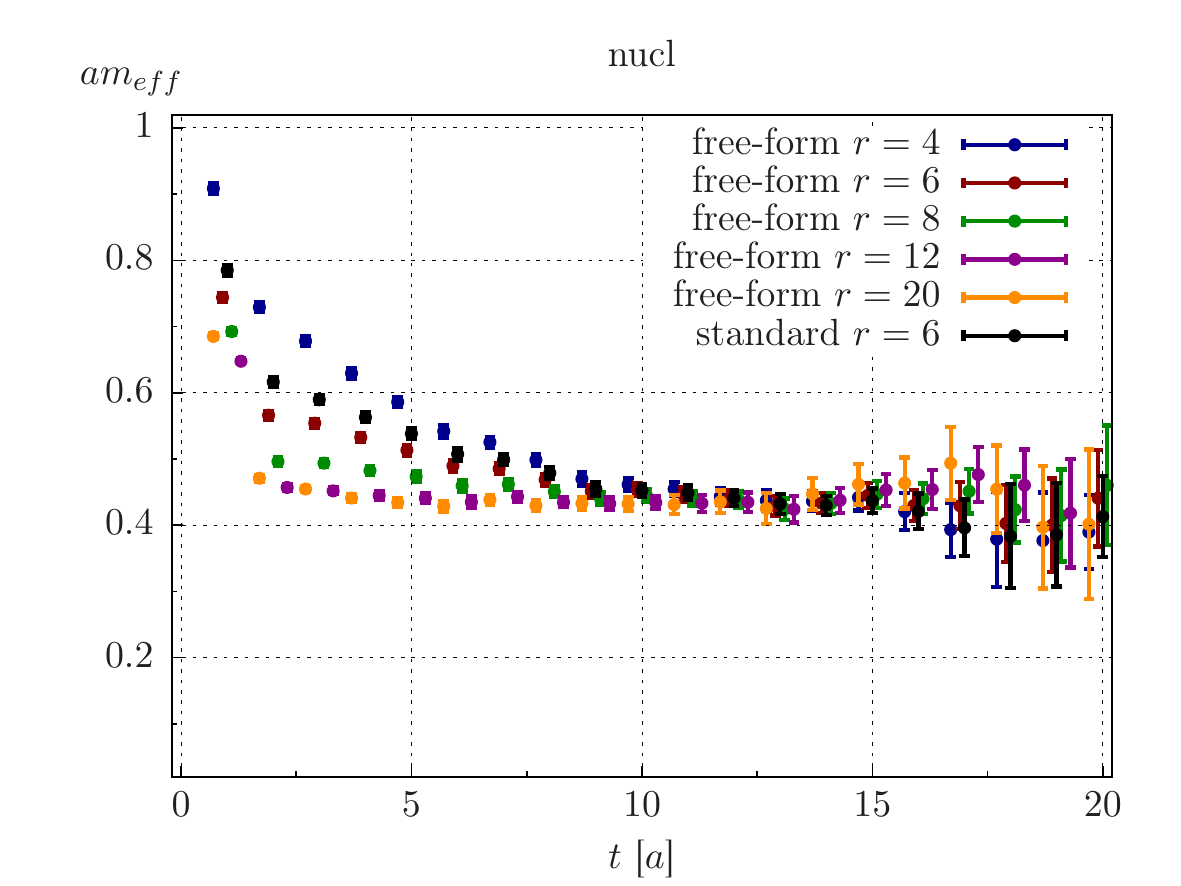}
\end{center}
\caption{Effective mass plots using free-form smeared Gaussian sources of varying
         radius for the pion (top) and the nucleon (bottom)
         on an $a=0.063$~fm ensemble with $N_{\rm f}=2$ flavours
         of O($a$)-improved Wilson fermions ($m_\pi\approx 450$~MeV,
         $N_{\rm cfg}=168$).
         Also shown for comparison is the corresponding result for the
         Gaussian smearing used to build the set of paths employed;
         note that the free-form smeared source of the same radius
         ($r=6a\approx 0.38$~fm) gives a slightly faster approach to the plateau.
         At very large smearing radii it can be seen that the procedure
         saturates in the sense that no further improvement is seen by going
         from $r=12a\approx 0.76$~fm to $r=20a\approx 1.3$~fm.}
\label{fig:effmassE5}
\end{figure}

\begin{figure}
\begin{center}
\includegraphics[width=0.69\textwidth,keepaspectratio=]{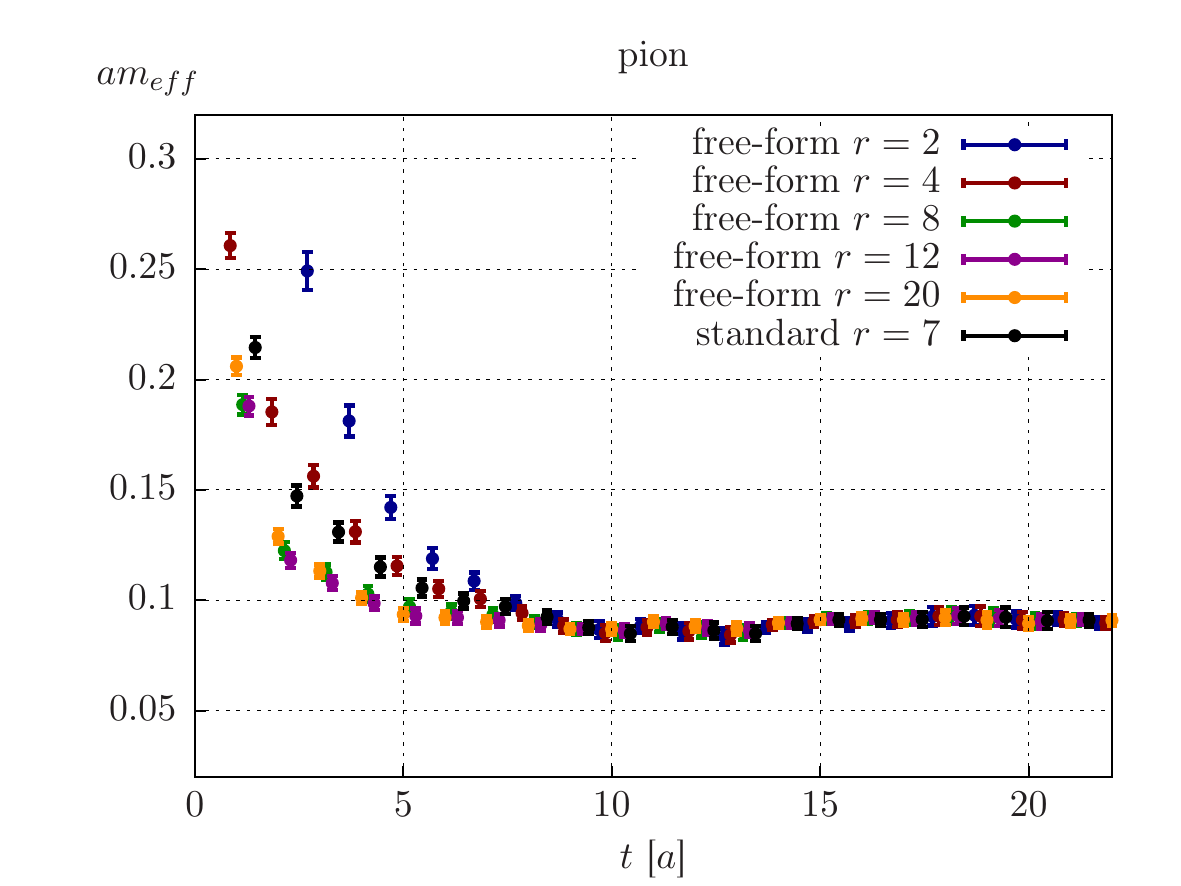}\\
\includegraphics[width=0.69\textwidth,keepaspectratio=]{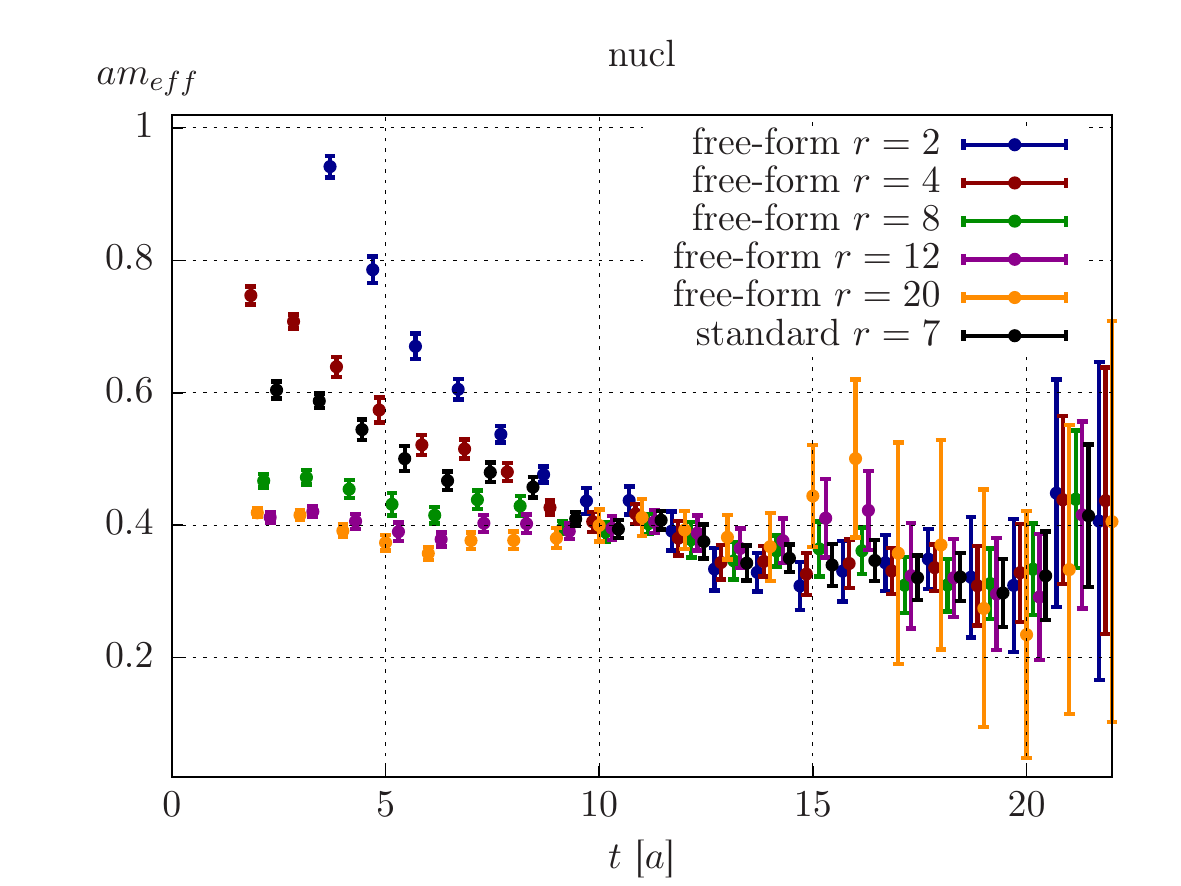}
\end{center}
\caption{Effective mass plots using free-form smeared Gaussian sources of varying
         radius for the pion (top) and the nucleon (bottom)
         on an $a=0.063$~fm ensemble with $N_{\rm f}=2$ flavours of
         O($a$)-improved Wilson fermions ($m_\pi\approx 280$~MeV,
         $N_{\rm cfg}=50$).
         Also shown for comparison is the corresponding result for the
         Gaussian smearing used to build the set of paths employed.
         Compared to figure~\ref{fig:effmassE5},
         this ensemble has a lighter pion, and we observe a quicker growth
         in the statistical errors at large euclidean times; this is
         particularly true for larger smearing radii, where, however, the earlier
         onset of the plateau behaviour more than compensates any resulting
         loss of statistical accuracy.}
\label{fig:effmassF7}
\end{figure}

\begin{figure}
\begin{center}
\includegraphics[width=0.69\textwidth,keepaspectratio=]{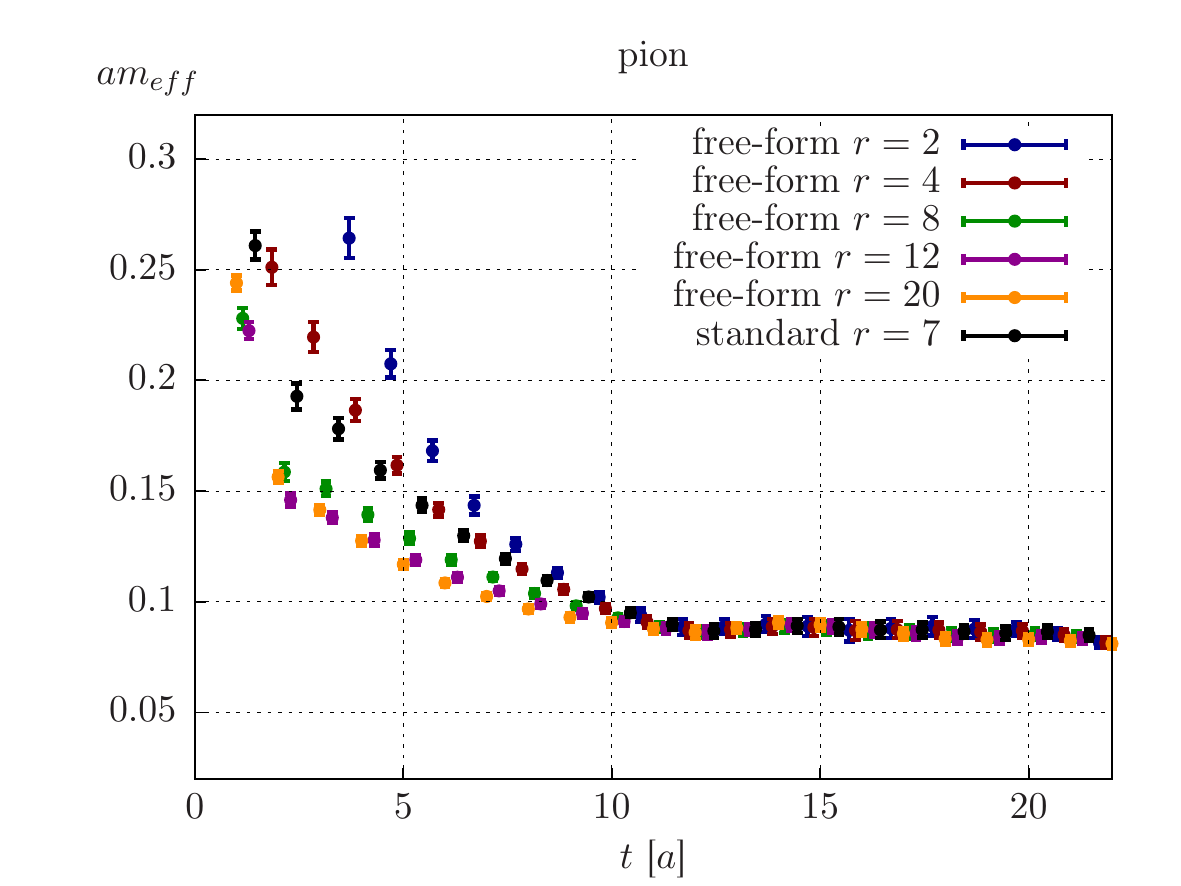}\\
\includegraphics[width=0.69\textwidth,keepaspectratio=]{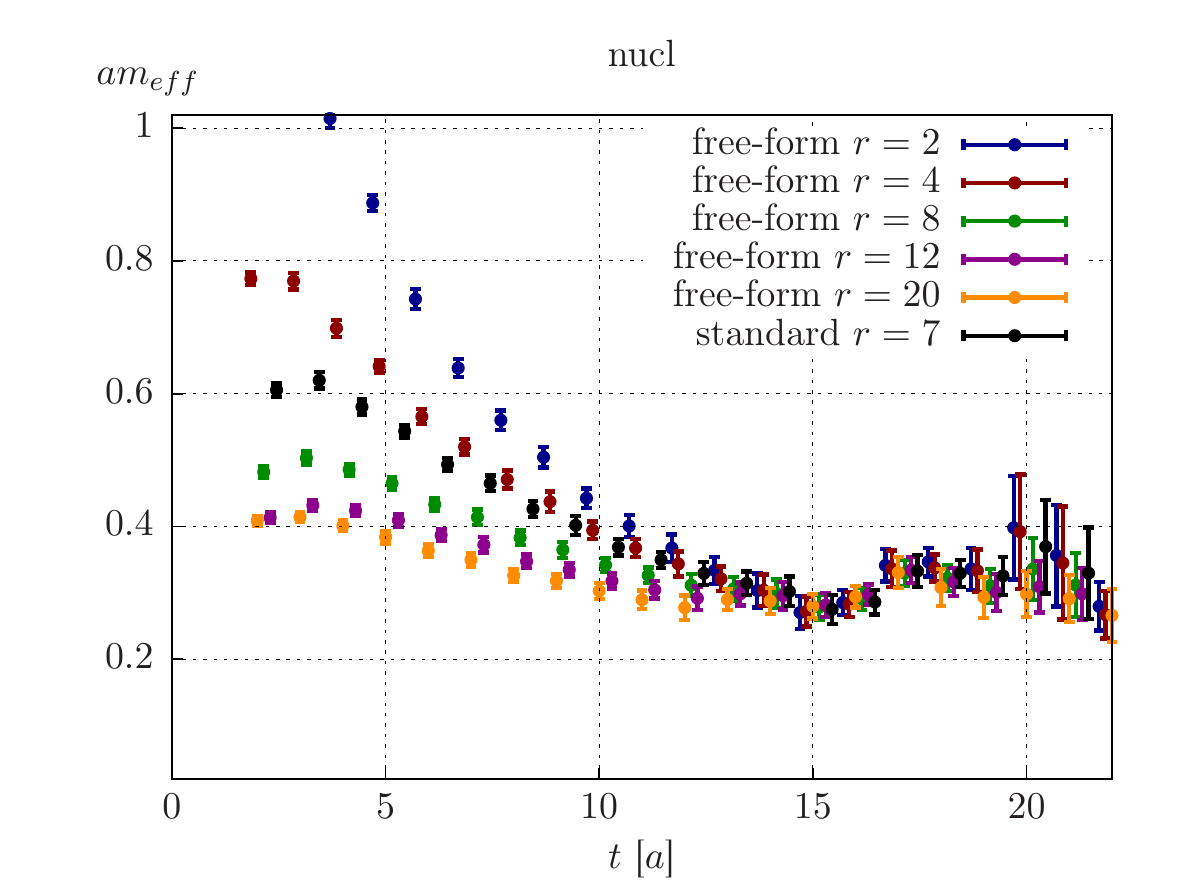}
\end{center}
\caption{Effective mass plots using free-form smeared Gaussian sources of varying
         radius for the pion (top) and the nucleon (bottom)
         on an $a=0.050$~fm ensemble with $N_{\rm f}=2$ flavours
         of O($a$)-improved Wilson fermions ($m_\pi\approx 320$~MeV,
         $N_{\rm cfg}=50$).
         Also shown for comparison is the corresponding result for the
         Gaussian smearing used to build the set of paths employed.
         This ensemble has a similar pion mass as
         the one in figure~\ref{fig:effmassE5}, but a finer lattice spacing;
         the efficiency of free-form smearing appears to be only mildly affected
         by the change in $a$.}
\label{fig:effmassN6}
\end{figure}

To compare the projection properties of free-form smeared Gaussians of various
radii with those of standard Gaussian smearing, we have computed the
corresponding hadron correlators on a range of ensembles%
\footnote{The ensembles used were generated within the
          Coordinated Lattice Simulations (CLS) initiative
          \cite{CLS}
          using the DD-HMC algorithm
          \cite{Luscher:2005rx,DDHMC}.
};
the effective masses for both the pion and nucleon channels on the various
ensembles can be seen in figures~\ref{fig:effmassE5} to \ref{fig:effmassN6}.
It is evident that free-form smearing allows us to use large radii,
thus strongly improving the overlap with the ground state. The effect is
particularly dramatic in the nucleon channel. As a consequence, a much earlier
onset of the plateau region is observed, which significantly improves the
statistical precision of the fitted plateau value for the mass,
while maintaining a good handle on the contaminations from excited states.
This is of particular importance in the nucleon sector, where the rapid
growth of the noise-to-signal ratio imposes a limit on how late one can begin
a plateau fit with reasonable statistical errors, whereas the systematic errors
incurred from excited-state contributions limit how early a reasonably safe
plateau can be found. Suppressing the unwanted excited-state contributions
thus allows for earlier, much more statistically accurate plateau fits.


\section{Discussion and Outlook}\label{sec.discussion}

In lattice simulations, the use interpolating operators with good projection
properties to the ground state is crucial, in particular in the baryon sector,
where the signal-to-noise ratio deteriorates rapidly with the Euclidean time
separation between source and sink, making an early onset of the ground state
plateau region highly desirable. Covariant smearing operations are commonly
used to improve the overlap with the ground state in correlation function.
Unfortunately, as $a\to 0$, iterative procedures rapidly require high iteration
numbers to realize the smearing radii associated with a good ground-state
overlap, and we have seen that the resulting sources bear no resemblance to
the Gaussian shapes one would expect.

This strong distortion of the shape of iteratively covariantly smeared sources
found at large iteration counts leads to a number of problems:
firstly, it becomes increasingly hard to reach large smearing radii, and
secondly, the heuristic motivation for smearing is lost:
it is not clear why one should expect such strongly distorted sources to
provide an improved overlap with the hadronic ground state.

Here, we have given an interpretation of the source shape distortion in terms
of interference between paths winding around regions of large chromomagnetic
flux on a given configuration in a manner reminiscent of a non-abelian analogue
of the Aharonov-Bohm effect, cf. eqn.~(\ref{eqn:smear:ab}).

To bypass the resulting limitations on the achievable smearing radii, and to
enable the use of covariantly smeared sources of arbitrary shape, we propose
a new smearing method, ``free-form smearing''. First tests
indicate that free-form smeared sources can reliably reach larger smearing radii
and yield better overlap with the ground state than can be achieved with the
more conventional methods of covariant smearing.

It might be expected that free-form smearing, besides improving the onset of the
plateau region for baryons, will be particularly useful for
applications in quarkonium spectroscopy, where the ability to create arbitrary
source shapes (such as hydrogenic wavefunctions) is important.
A notable limitation of the free-form smearing method is that in its current form
it only applies to smearing at the source, precluding its use in the variational
method, where the same operators are needed at both source and sink.
In order to apply free-form smearing at the sink, the rescaling function $N(x)$
would have to be computed with the original point source put at each point on
the sink timeslice, leading to a runtime that scales with $V^2$.
However, a number of methods
\cite{vonHippel:2007ar,Hornbostel:2011hu,Fleming:2009wb}
capable of accessing excited state information from a vector of correlation
functions exist, for which free-form smearing would appear to be highly suitable.

As a first practical application, we intend to apply free-form smearing to the
determination of the lattice spacing from the $\Omega$ mass, which we had
previously performed in
\cite{Capitani:2011fg};
the rapid approach to the plateau found for free-form smeared baryonic correlation
functions leads us to expect that a significant reduction of both statistical
and systematic errors should become possible with free-form smearing in this
context.


\acknowledgments

We acknowledge useful discussions with Rainer Sommer and Hubert Simma. \\
Our calculations were performed on the ``Wilson'' HPC Cluster
at the Institute for Nuclear Physics, University of Mainz.
We thank Dalibor Djukanovic and Christian Seiwerth for technical support.
We are grateful for computer time allocated to project HMZ21
on the BG/Q ``JUQUEEN'' computer at NIC, J\"ulich.
This work was granted access to the HPC resources
of the Gauss Center for Supercomputing at Forschungzentrum J\"ulich, Germany,
made available within the Distributed European Computing Initiative
by the PRACE-2IP, receiving funding from
the European Community's Seventh Framework Programme (FP7/2007-2013)
under grant agreement RI-283493.
This work was supported by the DFG
via SFB~1044
and grant HA~4470/3-1. \\
The 3d plots shown in this paper were generated using matplotlib
\cite{Hunter:2007matplotlib}. \\
We are grateful to our colleagues within the CLS initiative
for sharing ensembles.


\bibliographystyle{JHEP}
\bibliography{smearing}

\providecommand{\href}[2]{#2}\begingroup\raggedright\begin{thebibliography}{10}

\bibitem{Capitani:2010sg}
S.~Capitani, B.~Knippschild, M.~Della~Morte, and H.~Wittig, {\it {Systematic
  errors in extracting nucleon properties from lattice QCD}},  {\em PoS} {\bf
  LATTICE2010} (2010) 147, [\href{http://xxx.lanl.gov/abs/1011.1358}{{\tt
  1011.1358}}].

\bibitem{Dinter:2011sg}
S.~Dinter, C.~Alexandrou, M.~Constantinou, V.~Drach, K.~Jansen, {\em et~al.},
  {\it {Precision Study of Excited State Effects in Nucleon Matrix Elements}},
  {\em Phys.Lett.} {\bf B704} (2011) 89--93,
  [\href{http://xxx.lanl.gov/abs/1108.1076}{{\tt 1108.1076}}].

\bibitem{Alexandrou:2013xon}
C.~Alexandrou, S.~Dinter, V.~Drach, K.~Hadjiyiannakou, K.~Jansen, {\em et~al.},
  {\it {A Stochastic Method for Computing Hadronic Matrix Elements}},
  \href{http://xxx.lanl.gov/abs/1302.2608}{{\tt 1302.2608}}.

\bibitem{Green:2011fg}
J.~Green, J.~Negele, A.~Pochinsky, S.~Krieg, and S.~Syritsyn, {\it {Excited
  state contamination in nucleon structure calculations}},  {\em PoS} {\bf
  LATTICE2011} (2011) 157, [\href{http://xxx.lanl.gov/abs/1111.0255}{{\tt
  1111.0255}}].

\bibitem{Green:2012ud}
J.~Green, M.~Engelhardt, S.~Krieg, J.~Negele, A.~Pochinsky, {\em et~al.}, {\it
  {Nucleon Structure from Lattice QCD Using a Nearly Physical Pion Mass}},
  \href{http://xxx.lanl.gov/abs/1209.1687}{{\tt 1209.1687}}.

\bibitem{Capitani:2012gj}
S.~Capitani, M.~Della~Morte, G.~von Hippel, B.~J{\"a}ger, A.~J{\"u}ttner, {\em
  et~al.}, {\it {The nucleon axial charge from lattice QCD with controlled
  errors}},  {\em Phys.Rev.} {\bf D86} (2012) 074502,
  [\href{http://xxx.lanl.gov/abs/1205.0180}{{\tt 1205.0180}}].

\bibitem{Michael:1983gevp}
C.~Michael and I.~Teasdale, {\it {Extracting Glueball Masses from Lattice
  QCD}},  {\em Nucl.Phys.} {\bf B215} (1983) 433.

\bibitem{Luscher:1990ck}
M.~L{\"u}scher and U.~Wolff, {\it How to calculate the elastic scattering
  matrix in two-dimensional quantum field theories by numerical simulation},
  {\em Nucl.Phys.} {\bf B339} (1990) 222--252.

\bibitem{Blossier:2009kd}
{\bf ALPHA} Collaboration, B.~Blossier, M.~Della~Morte, G.~von Hippel,
  T.~Mendes, and R.~Sommer, {\it {On the generalized eigenvalue method for
  energies and matrix elements in lattice field theory}},  {\em JHEP} {\bf
  0904} (2009) 094, [\href{http://xxx.lanl.gov/abs/0902.1265}{{\tt
  0902.1265}}].

\bibitem{Gusken:1989ad}
S.~G{\"u}sken, U.~L{\"o}w, K.~H. M{\"u}tter, R.~Sommer, A.~Patel, {\em et~al.},
  {\it {Nonsinglet Axial Vector Couplings of the Baryon Octet in Lattice QCD}},
   {\em Phys.Lett.} {\bf B227} (1989) 266.

\bibitem{Alexandrou:1990dq}
C.~Alexandrou, F.~Jegerlehner, S.~G{\"u}sken, K.~Schilling, and R.~Sommer, {\it
  {B meson properties from lattice QCD}},  {\em Phys.Lett.} {\bf B256} (1991)
  60--67.

\bibitem{Allton:1993wc}
{\bf UKQCD} Collaboration, C.~Allton {\em et~al.}, {\it {Gauge invariant
  smearing and matrix correlators using Wilson fermions at Beta = 6.2}},  {\em
  Phys.Rev.} {\bf D47} (1993) 5128--5137,
  [\href{http://xxx.lanl.gov/abs/hep-lat/9303009}{{\tt hep-lat/9303009}}].

\bibitem{Basak:2005aq}
S.~Basak, R.~Edwards, G.~Fleming, U.~Heller, C.~Morningstar, {\em et~al.}, {\it
  {Group-theoretical construction of extended baryon operators in lattice
  QCD}},  {\em Phys.Rev.} {\bf D72} (2005) 094506,
  [\href{http://xxx.lanl.gov/abs/hep-lat/0506029}{{\tt hep-lat/0506029}}].

\bibitem{Basak:2005ir}
{\bf Lattice Hadron Physics Collaboration (LHPC)} Collaboration, S.~Basak {\em
  et~al.}, {\it {Clebsch-Gordan construction of lattice interpolating fields
  for excited baryons}},  {\em Phys.Rev.} {\bf D72} (2005) 074501,
  [\href{http://xxx.lanl.gov/abs/hep-lat/0508018}{{\tt hep-lat/0508018}}].

\bibitem{Hart:2009nr}
A.~Hart, G.~M. von Hippel, R.~R. Horgan, and E.~H. M{\"u}ller, {\it {Automated
  generation of lattice QCD Feynman rules}},  {\em Comput. Phys. Commun.} {\bf
  180} (2009) 2698--2716, [\href{http://xxx.lanl.gov/abs/0904.0375}{{\tt
  0904.0375}}].

\bibitem{DellaMorte:2012xc}
M.~Della~Morte, B.~J{\"a}ger, T.~Rae, and H.~Wittig, {\it {Improved
  interpolating fields for hadrons at non-zero momentum}},  {\em Eur.Phys.J.}
  {\bf A48} (2012) 139, [\href{http://xxx.lanl.gov/abs/1208.0189}{{\tt
  1208.0189}}].

\bibitem{Dudek:2007wv}
J.~J. Dudek, R.~G. Edwards, N.~Mathur, and D.~G. Richards, {\it {Charmonium
  excited state spectrum in lattice QCD}},  {\em Phys.Rev.} {\bf D77} (2008)
  034501, [\href{http://xxx.lanl.gov/abs/0707.4162}{{\tt 0707.4162}}].

\bibitem{vonHippel:2007ar}
G.~M. von Hippel, R.~Lewis, and R.~G. Petry, {\it {Evolutionary Fitting Methods
  for the Extraction of Mass Spectra in Lattice Field Theory}},  {\em
  Comput.Phys.Commun.} {\bf 178} (2008) 713--723,
  [\href{http://xxx.lanl.gov/abs/0707.2788}{{\tt 0707.2788}}].

\bibitem{Hornbostel:2011hu}
K.~Hornbostel, G.~Lepage, C.~Davies, R.~Dowdall, H.~Na, {\em et~al.}, {\it
  {Fast Fits for Lattice QCD Correlators}},  {\em Phys.Rev.} {\bf D85} (2012)
  031504, [\href{http://xxx.lanl.gov/abs/1111.1363}{{\tt 1111.1363}}].

\bibitem{Fleming:2009wb}
G.~T. Fleming, S.~D. Cohen, H.-W. Lin, and V.~Pereyra, {\it {Excited-State
  Effective Masses in Lattice QCD}},  {\em Phys.Rev.} {\bf D80} (2009) 074506,
  [\href{http://xxx.lanl.gov/abs/0903.2314}{{\tt 0903.2314}}].

\bibitem{Dowdall:2011wh}
{\bf HPQCD} Collaboration, R.~Dowdall {\em et~al.}, {\it {The Upsilon spectrum
  and the determination of the lattice spacing from lattice QCD including charm
  quarks in the sea}},  {\em Phys.Rev.} {\bf D85} (2012) 054509,
  [\href{http://xxx.lanl.gov/abs/1110.6887}{{\tt 1110.6887}}].

\bibitem{CLS}
{Coordinated Lattice Simulations (CLS)}.
\newblock \url{https://twiki.cern.ch/twiki/bin/view/CLS/WebIntro}.

\bibitem{Luscher:2005rx}
M.~L{\"u}scher, {\it {Schwarz-preconditioned HMC algorithm for two-flavour
  lattice QCD}},  {\em Comput.Phys.Commun.} {\bf 165} (2005) 199--220,
  [\href{http://xxx.lanl.gov/abs/hep-lat/0409106}{{\tt hep-lat/0409106}}].

\bibitem{DDHMC}
M.~L{\"u}scher, ``{DD-HMC algorithm for two-flavour lattice QCD}.''
\newblock \url{http://luscher.web.cern.ch/luscher/DD-HMC/index.html}.

\bibitem{Capitani:2011fg}
S.~Capitani, M.~Della~Morte, G.~von Hippel, B.~Knippschild, and H.~Wittig, {\it
  {Scale setting via the $\Omega$ baryon mass}},  {\em PoS} {\bf LATTICE2011}
  (2011) 145, [\href{http://xxx.lanl.gov/abs/1110.6365}{{\tt 1110.6365}}].

\bibitem{Hunter:2007matplotlib}
J.~D. Hunter, {\it Matplotlib: A 2d graphics environment},  {\em Computing In
  Science \& Engineering} {\bf 9} (2007), no.~3 90--95.

\end{thebibliography}\endgroup

\end{document}